\documentclass[nonacm,sigplan]{acmart}

\usepackage{algorithmic}
\usepackage{graphicx}
\usepackage{caption}
\usepackage{subcaption}
\usepackage{textcomp}
\usepackage{xcolor}
\usepackage{color,soul}
\usepackage{pifont}
\usepackage{tabularx}
\usepackage{amsmath}
\usepackage{notoccite}
\usepackage[ruled,linesnumbered]{algorithm2e}

\SetCommentSty{mycommfont}
\SetArgSty{textnormal}

\author{Yuting Wu*, Ziyu Wang*, Wei D. Lu } 
\thanks{*Equal Contribution; listing order determined by random dice rolling.}
\affiliation{%
  \institution{Department of Electrical Engineering and Computer Science, the University of Michigan, Ann Arbor.}
  \city{Ann Arbor} 
  \state{Michigan} 
  \country{USA}
  \postcode{48105}
}
\email{{wuyt,  ziwa,  wluee}@umich.edu}

\begin{document}

\title{PIM-GPT: A Hybrid Process-in-Memory Accelerator for Autoregressive Transformers}

\begin{abstract}
Decoder-only Transformer models such as GPT have demonstrated exceptional performance in text generation, by autoregressively predicting the next token. However, the efficacy of running GPT on current hardware systems is bounded by low compute-to-memory-ratio and high memory access. Process-in-memory (PIM) architectures can minimize off-chip data movement and utilize high internal bandwidth. They stand out as promising candidates for accelerating memory-bounded tasks such as GPT inference.

In this work, we propose a PIM accelerator, PIM-GPT, which achieves end-to-end acceleration of GPT inference with high performance and high energy efficiency. PIM-GPT leverages DRAM-based PIM designs for executing multiply-accumulate (MAC) operations directly in the DRAM chips, eliminating the need to move matrix data off-chip. Non-linear functions and data communication is supported by an application specific integrated chip (ASIC). At the software level, mapping schemes are designed to maximize data locality and computation parallelism by concatenating and partitioning matrices among DRAM channels and banks to utilize all available in-memory computation units. The efficiency of the PIM-GPT architecture is verified through circuit synthesis and an event-driven clock-cycle accurate simulator. Overall, PIM-GPT achieves 41$-$137$\times$, 631$-$1074$\times$ speedup and 123$-$383$\times$, 320$-$602$\times$ energy efficiency over GPU and CPU baseline on 8 GPT models with up to 1.4 billion parameters.
\end{abstract}

\maketitle 
\pagestyle{plain} 

\section{Introduction}
Attention-based Transformer models have revolutionized natural language processing (NLP) \cite{vaswani2017attention}. Transformer models including GPT and BERT have demonstrated superior performance in many NLP tasks such as text generation \cite{devlin2018bert}\cite{openai2023gpt}, text classification \cite{sun2019fine}\cite{chang2020taming}\cite{garg2020bae}, and machine translation \cite{wang2019learning}\cite{yao2020multimodal} compared to convolution neural networks (CNNs) or recurrent neural networks (RNNs). GPT in particular has attracted widespread public interest in text generation. GPT is a decoder-only Transformer model that generates context in an autoregressive manner by producing a single token at a time \cite{openai2023gpt}. The sequential processing feature of GPT, on the other hand, results in notable under-utilization of existing hardware such as the GPU, particularly for small batch inference tasks. 

\begin{figure}[t]
  \centering
  \includegraphics[width=\linewidth]{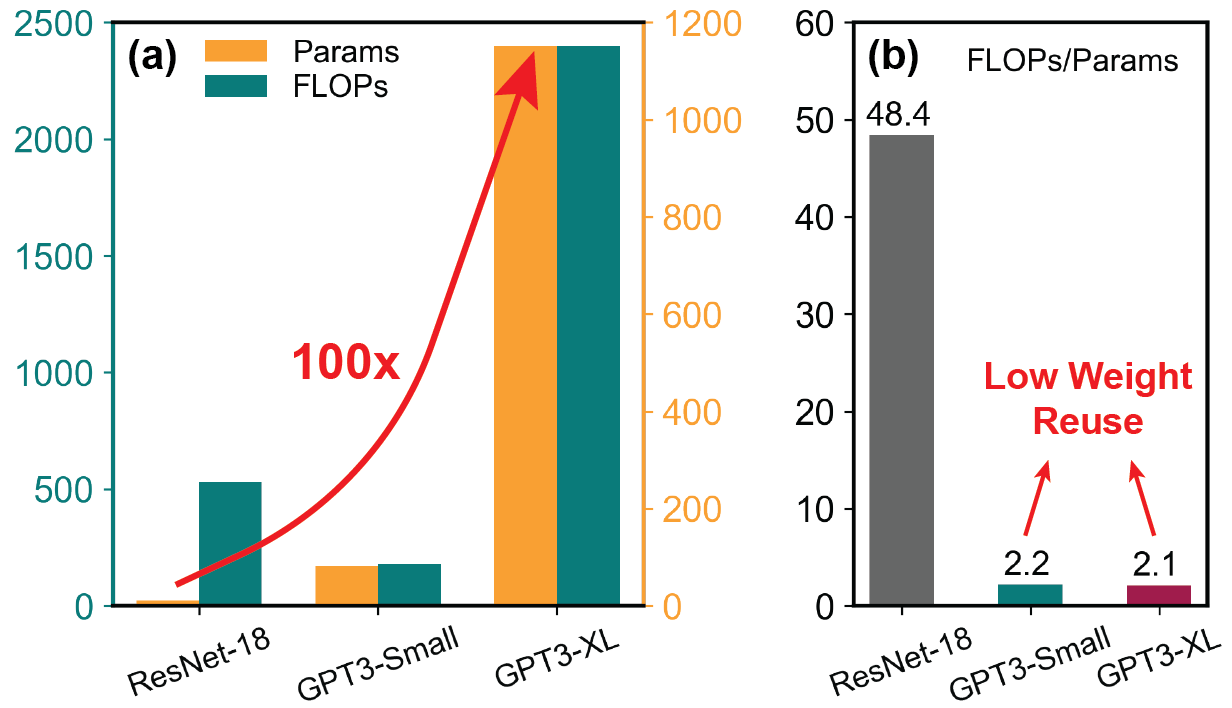}
  \caption{(a) Parameter and computation cost comparisons of GPTs and ResNet-18. (b) Operation/parameter ratios of CNN and GPT models.}
  \label{fig1}
\end{figure}

Compared to CNNs, GPT has two main features : (1) extremely large model size and (2) low compute-to-memory-ratio. As shown in Figure \ref{fig1}, the GPT3-XL model consists of 1.15 billion parameters \cite{brown2020language}, more than a hundred times higher than common CNNs such as ResNet-18 \cite{he2016deep}, while the arithmetic intensity per parameter (ops/parameter) is only 2.1, much lower than that of 48.4 in ResNet-18. These features mean large amounts of data have to be access through off-chip memory, while the massive amount of parallel computing units on GPU will be severely under-utilized, resulting in penalties in both performance and energy consumption. In the context of token generation in GPT, self-attention and the feed-forward networks (FFNs) rely on vector-matrix multiplication (VMM) instead of the more GPU-friendly matrix-matrix multiplication used for processing the entire sentences. In addition, these VMM operations are characterized by low data reuse as the weight values in the matrix are employed only once, in contrast to convolution operations where extensive reuse of the same weight is feasible.

Recently, several Transformer accelerators have been proposed to accelerate GPT inference \cite{wang2021spatten}\cite{hong2022dfx}\cite{zhou2022transpim}. However, these designs generally suffer from the following drawbacks: (1) expensive hardware overhead such as the usage of high-bandwidth memory (HBM) and dense in-memory logic; (2) model customization such as model and token pruning, which will make the architecture less flexible and cause accuracy loss; (3) lack of end-to-end acceleration as most studies only focus on the attention computation and feed-forward layers. Moreover, many existing Transformer accelerators are designed for encoder-only models like BERT, instead of decoder-only models like GPT \cite{ham20203}\cite{ham2021elsa}\cite{jang2019mnnfast}\cite{zadeh2020gobo}\cite{wang2023cta}\cite{dass2023vitality}\cite{you2023vitcod}.

DRAM-based process-in-memory (PIM) is a promising architecture to accelerate memory-bounded tasks \cite{mutlu2023memory}\cite{ghose2019processing}. The high storage capacity of DRAM allows all model parameters to be stored on chip. By integrating computation elements onto the DRAM chip, PIM allows the data to be consumed locally, utilizing high internal bandwidth and minimizing off-chip data movement. In principle, the placement of computation units for  multiply-accumulate (MAC) operations can be distributed across various regions, including sub-array, bank, I/O driver and logic die. Among these choices, placing MAC units at the bank level provides a favorable balance between performance, energy consumption and area cost \cite{shin2018mcdram}. Recent PIM developments indeed demonstrate efficient acceleration of MAC operations by integrating computation components at the bank level \cite{kwon202125}\cite{lee2021hardware}\cite{lee20221ynm}\cite{kwon20221ynm}. 

In this work, we propose PIM-GPT, a complete hardware-software solution for GPT inference acceleration. At the hardware level, PIM-GPT is a hybrid system that includes DRAM-based PIM chips to accelerate VMM near data and an application-specific integrated circuit (ASIC) to support other functions that are too expensive for PIM including necessary non-linear functions, data communication and instructions for the PIM chips. At the software level, mapping scheme is optimized to efficiently support the GPT dataflow. To accommodate the large model size and improve performance, the computation workloads are evenly distributed across available PIM channels and banks to maximize utilization of compute resources and on-chip bandwidth. Compared to existing Transformer accelerators \cite{wang2021spatten}\cite{hong2022dfx}\cite{zhou2022transpim}, the proposed PIM-GPT supports large GPT models end-to-end without the need of expensive HBM, making it an efficient and practical solution for GPT acceleration. Benchmarking analysis shows the proposed PIM-GPT achieves state-of-the-art speedup and energy efficiency for GPT inference tasks. The main contributions are as follows:
\begin{itemize}
    \item We design a hybrid system to accelerate GPT inference end-to-end using PIM with minimal changes to DRAM architecture, making it a practical solution. 
    \item We propose an optimized mapping scheme that maximizes data locality and distributes workloads across PIM channels and the ASIC to achieve high computation parallelism and efficiency. 
    \item We analyze performance, energy efficiency, and sensitivity of the complete system through circuit synthesis and a clock-cycle accurate simulator.
    \item PIM-GPT achieves state-of-the-art performance 41$-$\newline137$\times$, 631$-$1074$\times$ speedup and 123$-$383$\times$, 320$-$602$\times$ energy efficiency over GPU and CPU baseline, for 8 GPT models with up to 1.4 billion parameters.
\end{itemize}

\begin{figure}[t]
  \centering
  \includegraphics[width=\linewidth]{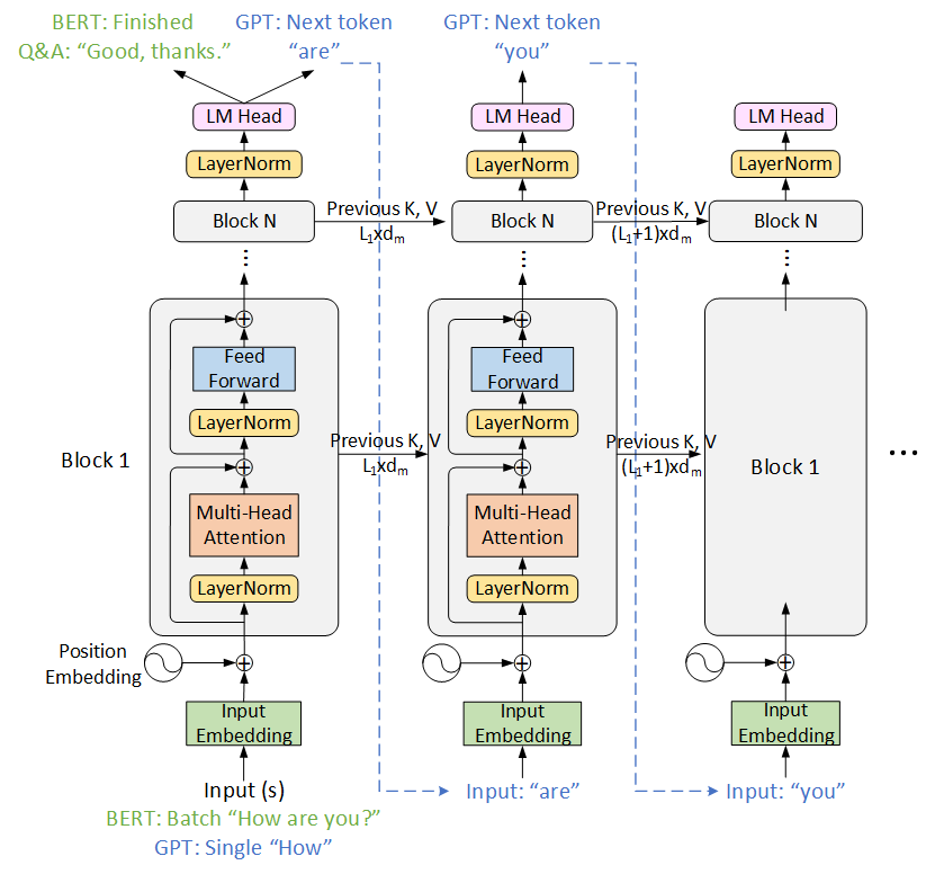}
  \caption{Transformer architectures of BERT and GPT.}
  \label{fig3}
\end{figure}

\section{Background and Motivation}
\subsection{Transformer Models}

Figure \ref{fig3} illustrates the typical structure of a Transformer model, which uses a self-attention mechanism that captures the relationship between different words in the sentence \cite{vaswani2017attention}. The original Transformer model consists of an encoder and a decoder, both containing \emph{N} identical transformer blocks. Each block includes a self-attention module and a FFN. Among them, BERT and GPT are the two most popular language models. Different from the encoder-only BERT that processes all input tokens at once \cite{devlin2018bert}, GPT is a decoder-only model which typically handles a single token at once and generates the next token in a sequential manner by attending to all previous tokens \cite{brown2020language}. Despite the differences, these two models share similar blocks, as shown in Figure \ref{fig3}.

For GPT, the input token is first transformed into a vector of dimension $d_{m}$ by the input embedding layer, where $d_{m}$ is the feature dimension of the model. The processed token is then fed into the Transformer blocks. The input token is first multiplied with three linear transformation matrices ($W_{K, Q, V}$) to obtain Query (q) , Key (k) and Value (v) vectors, where $ W_K \in\mathbb{R}^{d_{m}\times d_k}$, $W_Q \in\mathbb{R}^{d_{m }\times d_k}$, $W_V \in\mathbb{R}^{d_m\times d_v}$. The current Key and Value vectors k, v are then concatenated to the Key, Value matrices from the previous inputs. The q vector and Key, Value matrices are then passed to the self-attention heads to capture the dependencies between tokens with the following equation: 
\begin{equation}
    \text{Attention}\left(q,K,V\right) = \text{softmax}(\frac{q K^T}{\sqrt{d_k}})V
    \label{eq:1}
\end{equation}

\begin{figure*}[!t]
  \centering
  \includegraphics[width=\textwidth]{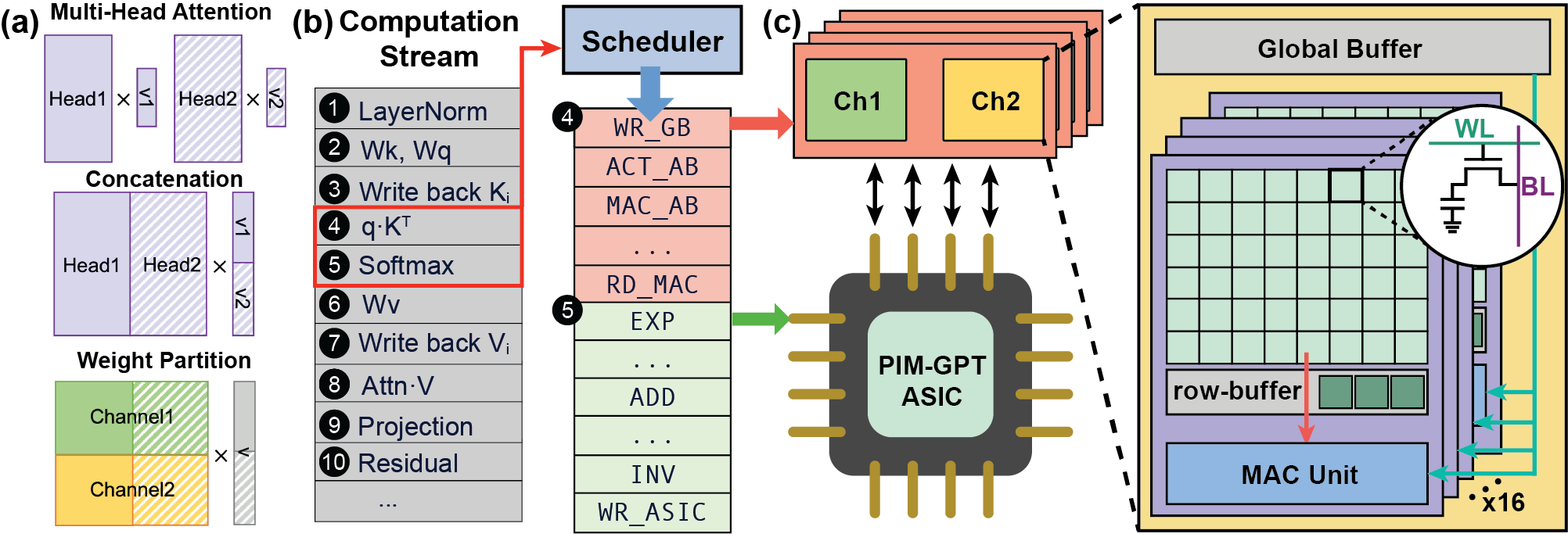}
  \caption{PIM-GPT system overview. (a) Hardware-aware GPT model partition. (b) Compilation of computation stream to command stream. (c) A complete PIM-GPT hardware architecture. 
}
  \label{fig4}
\end{figure*}

To allow the model to learn different relationships for each token, multi-head attention technique is adopted. The input vectors are split across attention heads, and each chunk goes through a separate head in parallel. All head outputs are combined by a linear projection layer to produce the final attention output. Following the multi-head attention, the attention output is fed into the FFN, which consists of two fully-connected layers with Gaussian Error Linear Unit (GELU) activation function \cite{hendrycks2016gaussian} in between. The output from the attention block is then applied as inputs to the next block for subsequent processing, repeating through N attention blocks. Afterwards, a final output layer predicts the next token. The content is generated autoregressively by repeating this process until reaches the required token length. 

BERT processes all input tokens in parallel and produces the outputs at once. Its core computation is matrix-matrix multiplication, and the performance is typically computation-bounded. In contrast, the core computation of autoregressive models such as GPT is the much lighter VMM in both multi-head attention and FFN layers. As a result,the arithmetic intensity of GPT is relatively low but the required memory access is high. As a result, compute-dense architectures such as GPUs are not efficient for GPT inference, while PIM techniques that leverage high internal bandwidth of DRAM chips are promising for GPT  acceleration. 

\subsection{DRAM-based PIM}
DRAM-based PIM architectures are promising candidates since DRAM is the mainstream memory in today's computing systems. The DRAM bank is a two-dimensional array of memory cells with a 1T1C structure, as shown in the inset of Figure \ref{fig4}(c). The cell represents the binary values of “1” or “0” by the presence or absence of charge. When reading data from a DRAM bank, the wordline will be activated and data stream from an entire row will be offloaded to the row-buffer. Employing computation units to directly access and consume data in the row-buffer is very efficient. However, the DRAM fabrication process is highly constrained. It only contains three metal layers that severely limit the complexity of the circuit, and the transistors are 3$\times$ slower than those in logic chips at the same node \cite{devaux2019true}\cite{gomez2022benchmarking}. Therefore, only limited logic and  buffers can be integrated on DRAM-based PIM.

DRAM-based PIM can be divided into (1) process-using-memory (PUM) and (2) process-near-memory (PNM). The first approach supports computing in DRAM banks or subarrays. It can yield higher level of parallelism and less data movement. However, to enable in-bank computation, extensive modifications to the subarray architecture are needed \cite{li2017drisa}\cite{gao2019computedram}\cite{jeong2022mac}. Additionally, only a limited instruction set is supported. The second approach implements computation logic near DRAM banks and can take advantage of the large internal bandwidth \cite{kwon202125}\cite{lee20221ynm}. By balancing the computation logic design and bank organization, PNM based PIM can achieve both high-performance and energy efficiency.

\subsection{Motivation}

Throughput-optimized processors like GPUs do not effectively accelerate autoregressive token generation due to the memory-bounded and sequential feature of decoders, particularly for inference tasks without batching. These architectures  suffer from intensive data transfer of large weight matrices, where the off-die data movement and DRAM interface have a significant energy consumption overhead \cite{keckler2011gpus}\cite{o2017fine}. Earlier PIM implementations focus on VMM acceleration in neural networks. But inter-layer functions in GPT such as LayerNorm and Residual can contribute $\sim10\%$ of total latency \cite{hong2022dfx}. Hence, efficient end-to-end PIM acceleration of GPT inference is of high interest.

PIM is a promising approach to relieve the memory bottleneck by storing matrix and performing VMM in memory. With weight matrices stationary in memory, only the input and output vectors will be transferred though the DRAM interface. Hence, the memory access complexity can be reduced from $\mathcal{O}(n^2)$ to $\mathcal{O}(n)$. Considering the high costs of adding logic to DRAM, it is more efficient to execute non-data intensive operations in a separate chip to achieve efficient end-to-end GPT acceleration. Besides hardware architecture advances, efficient workload distribution and dataflow management are required to fully exploit the high internal bandwidth of PIM. The proposed PIM-GPT is an end-to-end GPT accelerator with practical considerations in hardware implementation.

\begin{figure}[t]
  \centering
  \includegraphics[width=\linewidth]{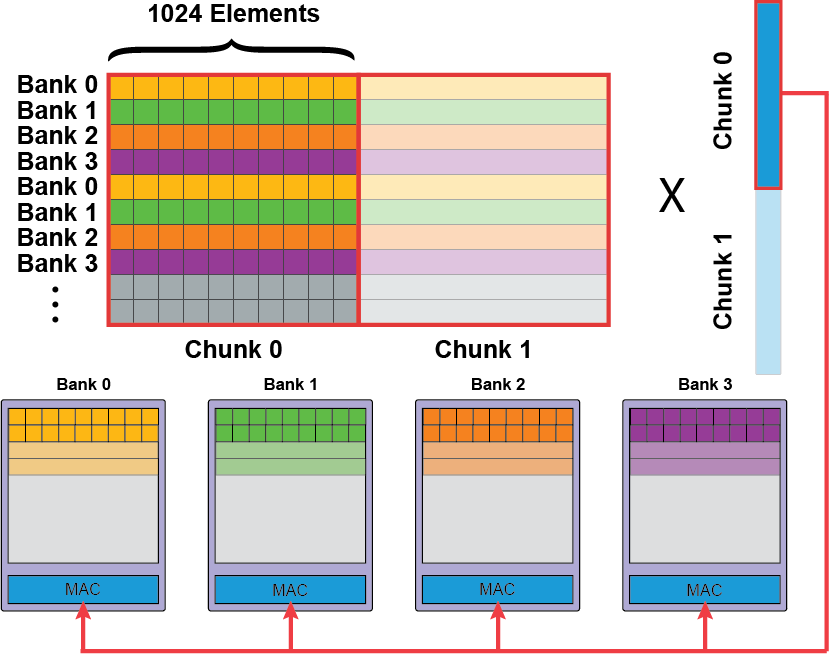}
  \caption{Multi-bank mapping scheme for VMM operation. Colors represent the different rows in the original matrix.}
  \label{fig2}
\end{figure}

\section{PIM-GPT Architecture}
\subsection{System Overview}
PIM-GPT is a memory-centric acceleration system aimed to support Transformer-based autoregressive token generation models including GPT. The PIM-GPT system is shown in Figure \ref{fig4}, which composes of PIM chips and a custom designed ASIC. The design principle of PIM is to maximally leverage data locality and parallelism to achieve high system performance and energy efficiency during VMM in attention and FFN. To achieve efficient VMM, PIM-GPT strategically partitions the model, taking hardware resources into account. It achieves parallel computing by broadcasting a single vector and reducing instruction overhead. PIM-GPT employs a specific mapping scheme for attention mechanism, as shown in Figure \ref{fig4}(a). Weight values from multiple attention heads are concatenated to accommodate the physical capacity of DRAM banks. The concatenated attention matrices, along with weights in FFN layers, are distributed to all channels and banks for parallel operation, following the mapping scheme in Figure \ref{fig2}, details of which will be elaborated in Section 4. PIM-GPT utilizes the mature PIM solution with a MAC unit integrated at a bank, which can be easily adopted from existing LPDDR4 \cite{shin2018mcdram} or GDDR6 \cite{lee20221ynm} design, making it a practical solution. The system integrates 8 channels, each of which executing VMM locally by broadcasting the vector from the buffer and loading matrices from DRAM bank arrays, as shown in Figure \ref{fig4}(c).

A high-level mapping scheme for VMM operation in PIM-GPT is shown in Figure \ref{fig2}. Since each row of the matrix is multiplied by the same vector, the rows are distributed across all banks. When the VMM begins, the vector is broadcasted to all banks and multiplied with matrix data from each bank in parallel. If the matrix dimension exceeds the physical storage of a bank row, it will be divided into chunks for mapping and computation. Similarly, multi-attention heads are concatenated to a large matrix to utilize parallel computing capability as well as maximize data locality. 

To facilitate end-to-end acceleration of large GPT models, non-VMM functions are executed on the ASIC chip. It is essential to highlight that PIM-GPT targets the elimination of off-chip movement of matrix data, requiring only the transfer of input/output vectors between PIM and ASIC for downstream computations, as well as data communication and intermediate data storage. This integration approach leverages the strengths of both PIM and ASIC, optimizing their capabilities to accelerate various computation tasks in the GPT computation stream with minimized data movement between them. Figure \ref{fig4}(b) illustrates the compilation of \ding{205}Attention and \ding{206}Softmax to command streams for the PIM and ASIC chips, respectively. All data in PIM-GPT are in bfloat16 (BF16) format, which preserves the approximate dynamic range of 32-bit floating point number to balance performance and accuracy.

\begin{figure}[t]
  \centering
  \includegraphics[width=\linewidth]{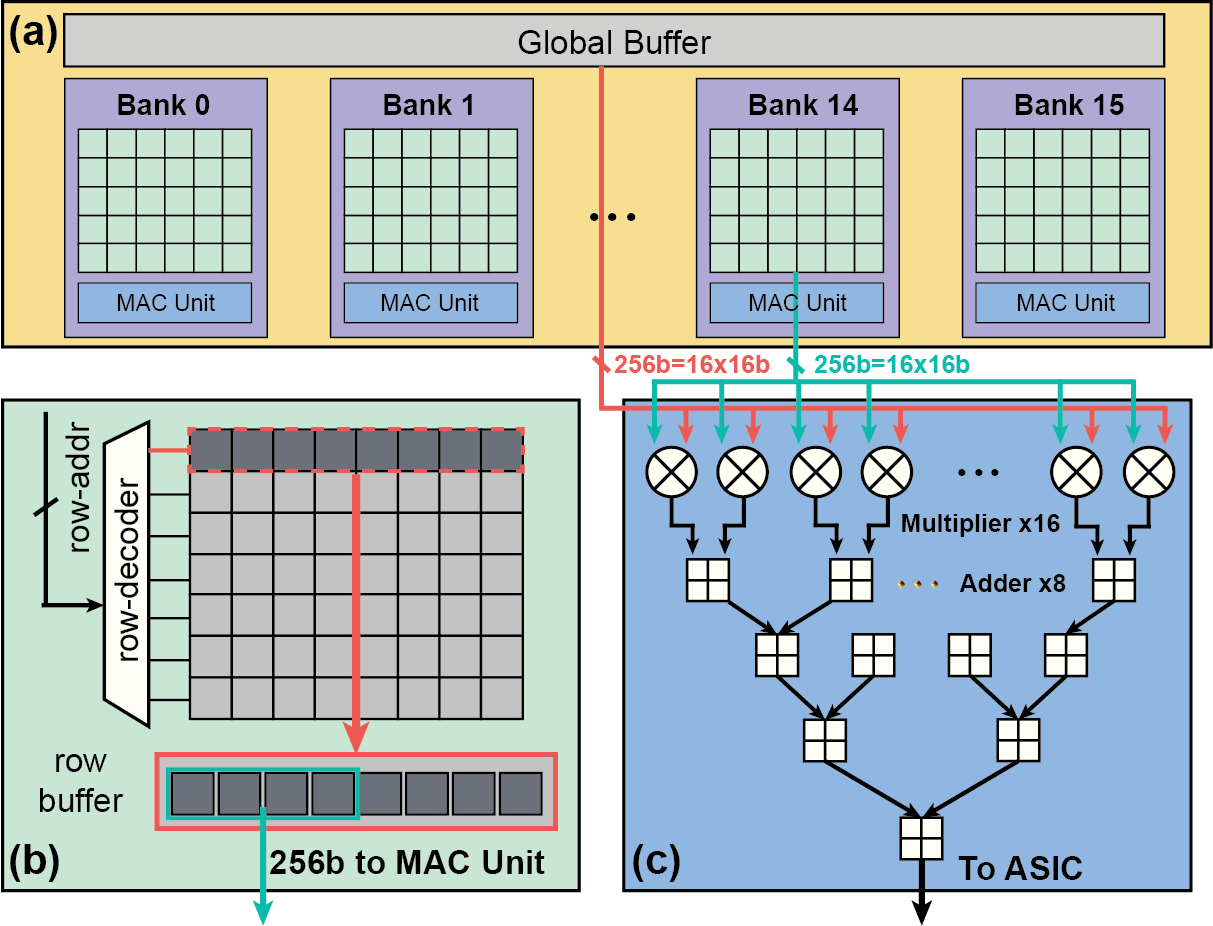}
  \caption{DRAM PIM organization. (a) A channel is composed of a global buffer and 16 banks. A bank contains (b) a conventional DRAM bank and (c) a MAC unit with multipliers and an adder tree.}
  \label{fig5}
\end{figure}

\subsection{DRAM-based PIM for VMM}
Placing MAC units and SRAM buffer inside DRAM modules has been widely used in PIM architectures for machine learning applications to minimize the memory access \cite{shin2018mcdram}\cite{kwon202125}\cite{lee20221ynm} \cite{he2020newton}. PIM-GPT places the MAC units at every bank. And all 16 banks can perform MAC simultaneously, as shown in Figure \ref{fig5}(a). Such MAC units arrangement has been proved feasible in both GDDR6 and HBM prototypes \cite{kwon202125}\cite{lee20221ynm}. Take GDDR6-based PIM as an example, placing MAC unit per bank offers 512~GB/s peak bandwidth per channel at 1~GHz (256 bits/bank $\times$ 16 banks). An SRAM buffer will be used to store and broadcast the input vector. However, integrating large SRAM buffers into DRAM is expensive and impractical. To provide a practical solution of GPT acceleration, we adopt 2~KB buffer size from \cite{lee20221ynm}, and broadcast vectors to all MAC units for parallel computing. When the input length exceeds the buffer size, vectors will be truncated into chunks, as shown in Figure \ref{fig2}. Partial MAC results will be computed and forwarded to the SRAM on the ASIC, followed by downstream partial sum execution on the ASIC. Compared to writing back to DRAM, forwarding the subportion of VMM results to ASIC reduces overall latency. Eliminating writing back also frees DRAM to perform subsequent VMMs. 

The bank organization is identical to conventional DRAM architectures, as shown in Figure \ref{fig5}(b). Once a row address is decoded, the entire corresponding row will be activated and all stored data will be forwarded to the row buffer. Herein, if the bank is not closed (precharged), data will be preserved in the row buffer. Reading data from the row buffer is much faster than from the bank array, since it skips the long latency row activation step. Hence, to maximize the utilization of peak internal bandwidth, data should be consumed from the row buffer as extensively as possible. This requires bank scheduling adheres to the open-row policy, wherein a row is not immediately precharged after data access. The goal is to employ this policy and optimally map the model to efficiently consume matrix data. In our  mapping scheme, matrix rows for VMM are partitioned and directly mapped to the same row addresses, thereby maximizing the row hit rate, as outlined at a high level in Figure \ref{fig2}.  

\begin{figure}[t]
  \centering
 \includegraphics[width=\linewidth]{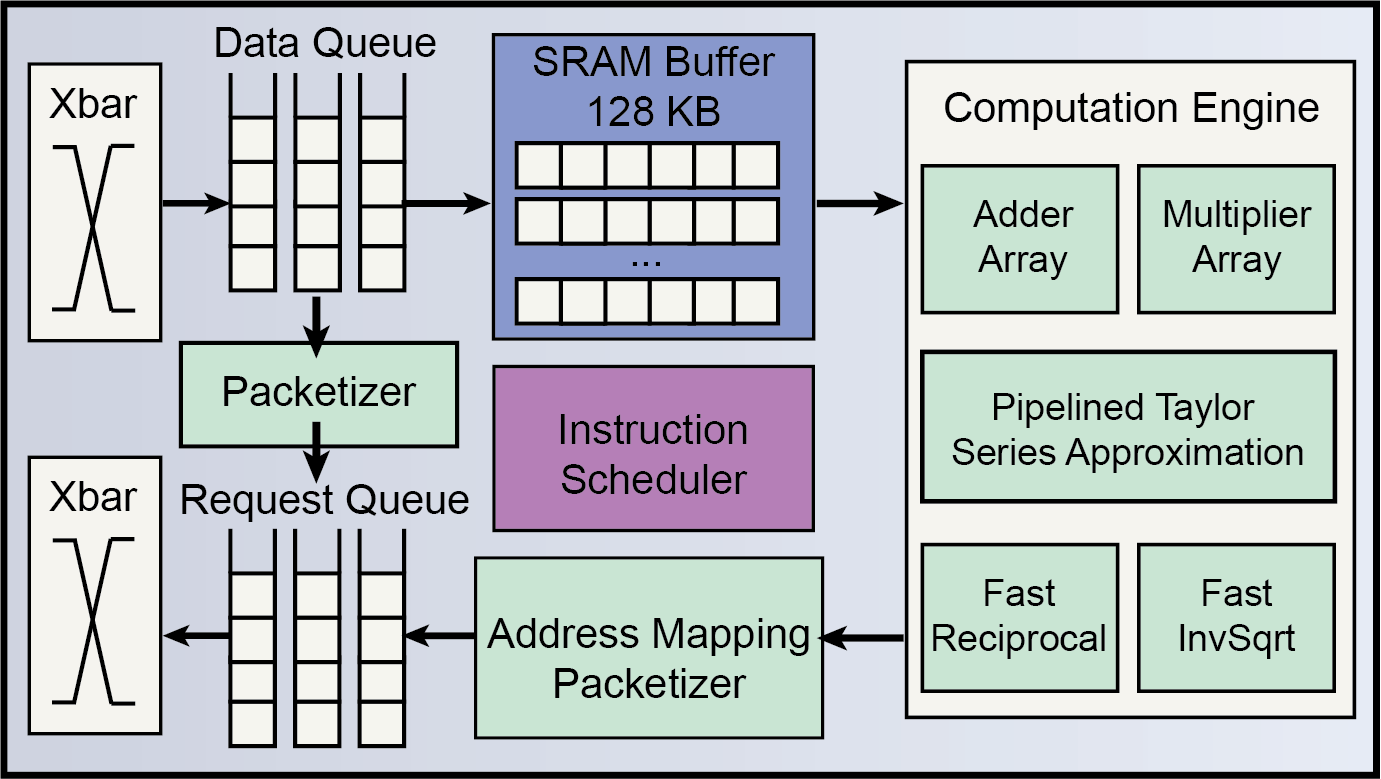}
  \caption{ASIC architecture of the PIM-GPT system.}
  \label{fig6}
\end{figure}

Inside each bank, only multipliers and an adder tree are implemented for MAC operation, as shown in Figure \ref{fig5}(c). Similar architecture has been proven effective in prior designs \cite{lee20221ynm}\cite{fan2023taskfusion}\cite{he2020newton}. During MAC operation, 16 vector values and the corresponding weights are fetched from the buffer and banks, respectively. The 16 multipliers multiply the vector data with weights. The adder tree accumulates the multiplication results for downstream computation. To minimize hardware cost and improve efficiency, PIM-GPT only performs VMM operations in the PIM while assigning all other computations, such as data bypassing, division and activation functions, to the ASIC. By doing so, the design allows the integration of lightweight MAC units to the DRAM chip to consume data locally without significantly sacrificing the memory capacity.

\subsection{ASIC Architecture}
The ASIC in PIM-GPT is used to support non-VMM arithmetic computations, manage data communication and intermediate data storage. The ASIC architecture is shown in Figure \ref{fig6}. Since VMM operations are performed in the PIM channels, data will only be read from DRAM when a VMM operation is done and requires downstream operations. The PIM channels communicate with the ASIC through memory bus and crossbar interconnects. The interconnects support data fetching from any DRAM channel and sending memory requests to a single channel or broadcasting to all channels after packeting the data with address. Data read from PIM have two possible paths on the ASIC: (1) writing back to banks in other PIM channels, such as Key, Value matrices for subsequent VMM, and (2) going through computation blocks in the ASIC, such as LayerNorm, Softmax, etc. 

If the data require downstream computation, they will be temporarily stored in the on-chip SRAM. The computation engines on ASIC are responsible for computations that cannot be run as MACs, such as non-linear activation functions, layer normalization, softmax and partial sum. The adders and multipliers in the computation engines follow the standard floating-point unit design to support summation and multiplication. For design reuse and performance considerations, other computation tasks are implemented with approximation algorithms using only addition and multiplication to achieve the required precision.

Three functions that require approximation are: (1) Softmax $s(x_i) = \frac{e^{x_i}}{\Sigma_{j=1}^{N}e^j}$; (2) LayerNorm $y = \frac{x - E[x]}{\sqrt{Var[x] + \epsilon}} \times \gamma + \beta$; (3)GELU$(x) = \frac{x}{2} \times [1 + erf(\frac{x}{\sqrt{2}})]$, which can be approximated by $\frac{x}{2} \times [1 + \tanh{\sqrt{2 / \pi}(x  + 0.044715 \times x^3)}]$.

The nonlinear function $e^x$, $\tanh{x}$, division and square root in these functions cannot be naively computed using addition and multiplication. Under given precision and data range, they can be efficiently approximated and converge in rapid iterations. Here $e^x$ and $\tanh{x}$ are computed using Taylor series approximation with the first six items, which can be computed with addition and multiplication.  

\begin{algorithm}[t]
\caption{Newton-Raphson Division}\label{alg:one}
\KwData{\textbf{\texttt{BFloat16}} $D = (S) M \times 2^E$}
\KwResult{\textbf{\texttt{BFloat16}} $\frac{1}{D}$}
\tcc{Scale by exponent subtraction}
$D' = D / 2^{(E+1)}$  \\
$X = \frac{48}{17} - \frac{32}{17} \times D'$ \\
\For{\textit{iter} \text{in} $\lceil \log_{2} \frac{P+1}{\log_{2} 17} \rceil$}{
$X = X + X \times (1 - D' \times X)$\\
}
\tcc{Scale the result}
$X' = X / 2^{(E+1)}$  \\
\end{algorithm}

The division operation is computed by multiplying the numerator with the inverse of the denominator. Both the reciprocal and inverse square root operations can be calculated with addition and multiplication following Newton’s method. A proper initial value is required to ensure the fast convergence. For reciprocal, we take advantage of Newton-Raphson division in Algorithm \ref{alg:one}. The fast inverse square root algorithm is adopted from Quake III Arena’s source code \cite{invsqrt}, as shown in Algorithm \ref{alg:two}.

The Newton-Raphson Division algorithm shown in Algorithm \ref{alg:one} is friendly for floating point numbers since it requires the input D to scale to a value close to 0, which can be easily done with exponent subtraction and mantissa shift in floating point data format. The P in line 3 is the precision of P binary places. Hence, for a 16-bit floating point number, it will take three iterations to get an accurate result. The fast inverse square root algorithm in Algorithm \ref{alg:two} unpacks the BF16 data into 16-bit integer data and padding 16-bit zeros to get an accurate approximation, followed by shift and subtraction from a constant. The pack step utilizes the 16 high bits of INT32 $L’$ to assemble a BF16 data’s sign bit, exponent and mantissa. In the fast square root algorithm, it can converge in a single step iteration. Here we take a conservative two step iteration.

\begin{algorithm}[t]
\caption{Fast Inverse Square Root}\label{alg:two}
\KwData{\textbf{\texttt{BFloat16}} $D = (S) M \times 2^E$}
\KwResult{\textbf{\texttt{BFloat16}} $\frac{1}{\sqrt{D}}$}
$D' = D \times 0.5$  \\
\tcc{Unpack data and pad with 0}
\textbf{\texttt{uint32\_t}} $L$ $\gets$ \{unpack($D'$), $0$x$0000$\} \\
$L' = 0$x5f3759df $- L >> 1$ \\
\tcc{Keep 16 high bits}
\textbf{\texttt{BFloat16}} $X$ $\gets$ pack($L'$)[$31:16$] \\
\For{\textit{iter} \text{in} IterNum}{
$X = X \times (1.5 - D' \times X \times X)$\\
}
\end{algorithm}

\section{PIM-GPT Dataflow}
\subsection{Overview}
PIM-GPT aims to distribute workloads among all PIM channels and ASIC efficiently. For VMM operation, the input vector is forwarded from the ASIC buffer to the SRAM buffer of all PIM channels, followed by broadcasting to all MAC units, as depicted in Figure \ref{fig7}. All MAC units will execute MAC of the same vector on different matrix partitions in parallel to fully utilize the PIM computation resources without instruction overhead. PIM-GPT implements the following techniques to coordinate workload between the PIM channels and the ASIC: (1) The partial outputs of VMM can be forwarded to the ASIC before the whole computation is completed, which effectively eliminates the data write back to DRAM banks; (2) When input vector length exceeds the buffer size, SRAM buffer on the ASIC are reserved to store intermediate data and the ASIC will accumulate partial VMM results from DRAM; (3) Pipelining between data transmission and computation, i.e. the ASIC will start operations on partially received vector while the rest are in transmission.

Mapping a model includes storing the weights to the allocated banks, as well as reserving space for the intermediate data (Key, Value matrices) for attention computation since they are dynamically expanded with token generation. To enhance the system performance, the mapping scheme is optimized to: (1) maximize row hit rate by exploiting data locality; (2) increase computational parallelism by balancing the workload across DRAM banks; and (3) reduce latency by minimizing data movement. At runtime, the system computes the bank address in the reserved space to write back Key and Value vectors. The high-level description of the mapping scheme is shown in Algorithm \ref{alg:three}. 

\subsection{Mapping Methods}
\noindent\underline{\textit{\textbf{Attention Head Mapping}}}: 

\begin{figure}[t]
  \centering
  \includegraphics[width=\linewidth]{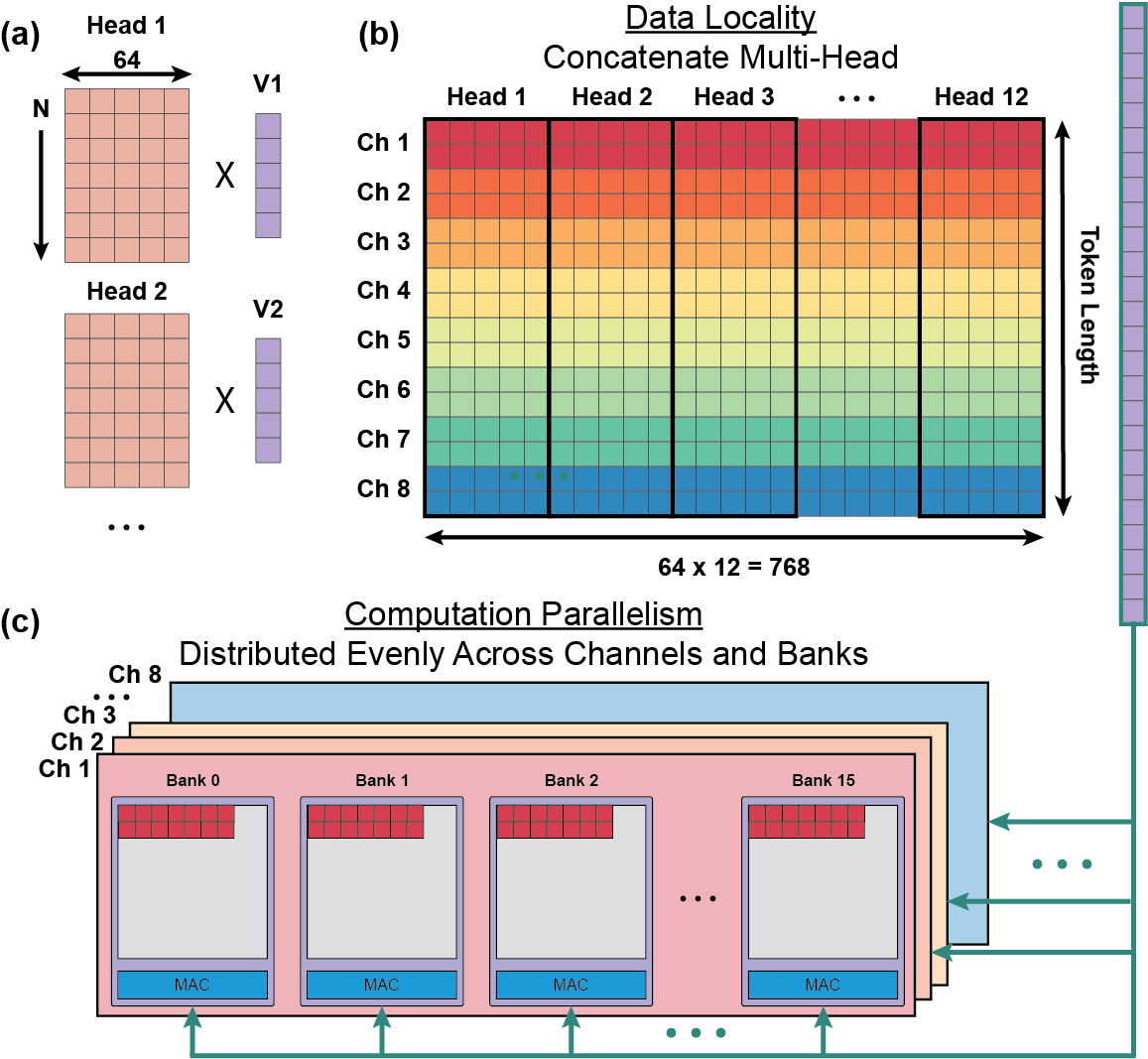}
  \caption{Mapping strategy for the Key Matrix. (a) Multi-head attention. (b) Concatenating multi-head to exploit data locality. (c) Distributing weight matrix evenly across channels and banks to maximize computation parallelism.}
  \label{fig7}
\end{figure}

As shown in Figure \ref{fig7}, the mapping scheme leverages data locality and maximizes computation parallelism.  Since activation (ACT) and precharge (PRE) commands are expensive in both latency and energy, achieving a high row hit rate is preferred. To this end, matrix data used for MAC operations need to be mapped to consecutive physical DRAM cells. This approach means the corresponding row only needs to be activated once to transfer all required data to the row buffer, and the MAC units can keep consuming data already in the opened row to minimize ACT and PRE operations. 

To take advantage of the data locality, it is desired that a row is fully mapped with data. However, a single attention head can be much smaller than the DRAM array dimension. As shown in Figure \ref{fig7}(a), attention head width of GPT2-small is 64 while a bank row can store 1024 16-bit data. To maximize the row hit rate, all attention heads in the same layer are concatenated to fill up the DRAM bank. Take GPT2-small as an example, 12 attention heads are concatenated to form a wider matrix with width of 768, along with the concatenation of input vectors, as shown in Figure \ref{fig7}(b). To maximize the utilization of MAC units, rows of the matrix are evenly distributed across PIM channels and banks. Figure \ref{fig7}(c) is a detailed example showing how the $K$ matrix are mapped through 8 PIM channels, assuming the token length is 256. First, attention heads in a layer are concatenated along column direction to form a larger matrix, with a dimension of $256 \times 768$. The concatenated matrix is mapped following the row major approach and evenly distributed as 32 matrix rows to each channel, as indicated by the rainbow colors in Figure \ref{fig7}(b) and (c). Inside each channel, all 16 banks are mapped with 2 rows of the matrix and execute MAC operations with the same vector in parallel.

\begin{algorithm}[t]
\caption{Model Mapping to PIM Banks}\label{alg:three}
\KwData{Computation Graph; PIM Configuration}
\KwResult{Memory Mapping and Reservation}
\tcc{Map weights to PIM banks}
\For{\textit{vmmBlock} \text{in} \textit{Computation Graph}}{
\If{vmmBlock.multiHead}{
hitScore $\gets$ maxRowHit(nhead, ncol)\\
vmmBlock $\gets$ concat(hitScore, vmmBlock)}
Mapping $\gets$ maxParallel(vmmBlock, nch, nbank)\\
}
\tcc{Reserve PIM bank rows for kv}
\For{\textit{wrBlock} \text{in} \textit{Computation Graph}}{
\If{block == write\_k}{
    hitScore $\gets$ maxRowHit(nhead, ncol, ltoken)\\
    wrBlock $\gets$ concat(hitScore, wrBlock)
    }
    Mapping $\gets$ Reserve(wrBlock, ltoken, nch, nbank)
}
\end{algorithm}

\noindent\underline{\textit{\textbf{Intermediate Data Memory Reservation}}}: 

Key and Value results need to be written back to the PIM banks and append to the existing Key and Value matrices. In the mapping stage, PIM-GPT reserves the required space in PIM banks for these intermediate data. Key and Value write-back are in row-major and column-major, respectively, since the transpose of Key matrix is required in Equation \ref{eq:1}, while not for Value matrix. 

PIM-GPT exploits data locality during write. During a token generation, a Key vector is produced by multi-attention heads, which is corresponding to $N=1$ in Figure \ref{fig7}(a). The Key vectors with length of 64 from 12 heads are concatenated to form a vector with length of 768, and written to the corresponding bank row reserved for the current token. The concatenated Key vectors produced by all token generation steps are evenly stored across all channels and banks for parallel downstream VMM computation. Value results are stored in column-major fashion because transpose is not required. To maximize the computation parallelism in the subsequent VMM, we distribute the Value matrix to all channels and banks using the same mapping scheme for Key matrix mapping as shown in Figure \ref{fig7}(c).

\noindent\underline{\textit{\textbf{Weight Matrix Tiling}}}: 

For larger GPT models, widths of both concatenated multi-head matrices and weight matrices in the FFN layers can exceed the capacity of a bank row. In this circumstance, the input vector length also exceeds the buffer size. Hence, both the matrix and the vector need to be truncated for VMM operation, as shown in Figure \ref{fig2}. The partial VMM results are then accumulated in the ASIC to facilite pipelined operation. Taking the input length of 2048 in Figure \ref{fig2} as an example, the matrix and the vector are sliced into two chunks. The elements in the same row will always be mapped to the same bank, as indicated by the faded color. Computation on the second chunk starts after the first chunk is completed to avoid frequently overwriting the SRAM buffer. The weight matrix tiling scheme distributes matrices evenly to achieve the highest possible DRAM channel-wise and bank-wise MAC computation parallelism.

\section{System Evaluation}
\subsection{Evaluation Method}
\textbf{Hardware Configuration}
To perform realistic estimates, we use GDDR6-based PIM prototype reported from SK Hynix for performance evaluation \cite{lee20221ynm}. The design only integrates lightweight MAC units near banks. The area of one processing unit (PU) is 0.19 $mm^{2}$ \cite{lee20221ynm}. All logic components of the ASIC are synthesized with SystemVerilog using Synopsys Design Compiler at TSMC 28nm HPC+ process node. Area and power of the logic components are obtained from the synthesis results. The area and power of SRAM buffer are extracted from the TSMC 28nm datasheet based on Synopsys Memory Compiler. The ASIC only consumes a core area of 0.64 $mm^{2}$, and the peak power is 304.59 $mW$. 

\begin{table}[h]
  \renewcommand*{\arraystretch}{1.4}
  \setlength{\arrayrulewidth}{0.25mm}
  \centering
  \caption{PIM-GPT Baseline Hardware Configuration.}
  \label{table:formatting}
  \begin{tabularx}{\linewidth}{|p{0.22\linewidth}|X|}
    \hline
    \textbf{Timing Constraint} & tRCD=12ns, tRP=12ns, tCCD=1ns, tWR=12ns, tRFC=455ns, tREFI=6825ns \\
    \hline
    \textbf{IDD} & IDD2N=276mA, IDD3N=262mA, IDD0=366mA, IDD4R=1590mA, IDD4W=1410mA, IDD5B=831mA \\
    \hline
    \textbf{GDDR6 \newline Specification} & Channel = 8, Banks/channel = 16, Capacity/channel = 4Gb, Row size = 2KB, Column number = 16k, Frequency = 1GHz, Pins = 16/channel, Data rate = 16Gb/s/pin \\
    \hline
    \textbf{PIM} & Buffer = 2KB/channel, MAC unit = 1/bank, Frequency = 1GHz, Power = 149.29mW \\
    \hline
    \textbf{AISC} & Frequency = 1GHz, SRAM = 128KB, \# of Adders = 256, \# of Multipliers = 128, Area = 0.64mm$^2$, Power = 304.59mW \\
    \hline
  \end{tabularx}
  \label{tab:1}
\end{table}

For DRAM energy benchmark analysis, we synthesized the optimized floating point MAC units at 28~nm technology, followed by scaling the voltage to 1.25V to match the GDDR6 supply voltage \cite{lee20221ynm}. Since routing is more complex in DRAM due to the limited metal layers compared to CMOS logic process, we conservatively multiply the power by 1.5, which comes to 149.29 $mW$ for 16 MAC units. As GDDR6 can be fabricated in 1ynm technology \cite{lee20221ynm}, the actual power consumption is expected to be lower than this estimate. Table \ref{tab:1} lists the timing constraints and current values used to model the PIM behavior for each command. For PIM related commands, the timing constraints are obtained from \cite{lee20221ynm}. For normal DRAM commands, we adopt GDDR5 timing constrains in \cite{kim2015ramulator} to make a conservative estimation due to the lack of detailed information of GDDR6. Similarly, the current values are obtained from DDR5 datasheet \cite{micron} and multiplied by 3 to account for the current consumption increase during parallel bank operations \cite{kwon20221ynm}. The GDDR6 I/O access energy 5.5~pJ/bit is adopted from GDDR6 datasheet \cite{gddr6}. The system performance in latency and power efficiency is evaluated based on these conservative assumptions. Detailed hardware configuration is summarized in Table \ref{tab:1}.

\textbf{Simulation Configuration} To evaluate the PIM-GPT system performance, we developed an event-driven clock-cycle accurate simulator in C++ that models the system behavior at token generation runtime. The simulator takes the GPT model and system configuration as inputs for model mapping. The computation graph is compiled into an instruction sequence following the hardware constraints.

The PIM and ASIC behaviors are modeled as state machines. The PIM hierarchy is organized as a tree structure at package, channel and bank levels. The PIM package node refers to the entire PIM portion that contains 8 channels as child nodes. Each channel consists of 16 banks as leaves. The state update will traverse down the tree from the root to the leaves. The same organization hierarchy can be found in other memory simulators, such as Ramulator \cite{kim2015ramulator}.       

The transition of states follows the timing constraints. At every clock cycle, the simulator checks the status of the ASIC and the PIM package. If both are in \texttt{Idle} state, the current instruction is completely consumed. It then fetches the next instruction, which will be decoded into command sequences. The ASIC chip or the PIM chip will be put into \texttt{Process} state after the instruction is issued. The simulator will compute the time \texttt{next\_time} that the ASIC or relevant PIM banks will take to complete the triggered events based on the latency model. The simulator keeps track of the status of all hardware components. If the \texttt{CLK} reaches the \texttt{next\_time}, the status of the corresponding node will be changed back to \texttt{Idle}.

\textbf{Benchmark Analysis}
The performance and energy efficiency of the proposed PIM-GPT system are evaluated using the simulator and compared to GPU (NVIDIA T4) and CPU (Intel Xeon Gold 6154 with 16Gb DDR4). 4 GPT2 \cite{radford2019language} and 4 GPT3 \cite{brown2020language} models with up to 1.4 billion parameters are used for the single batch inference benchmark analysis. DRAM energy is evaluated by multiplying the IDD values consumed during each command with the corresponding latency and VDD, following the standard procedure \cite{micron}\cite{ghose2018your}. DRAM refresh operations are also included. The energy consumed by the PIM MAC units and by the ASIC are computed by multiplying the latency reported by the simulator with the synthesized power consumption. The energy consumed through I/O is evaluated by the number of data transferred between the PIM chip and the ASIC chip. 

We select NVIDIA T4 as the GPU benchmark as it also uses GDDR6 as memory for a fair comparison. For GPU, latency is recorded using \texttt{torch.cuda.Event()}, and power is measured with \texttt{pynvml}, which is a wrapper around the NVIDIA management library. For CPU characterization, we use python package \texttt{time.time()} for latency measurement and an open-source terminal tool \texttt{s-tui} for power monitor. In each measurement, we generate 1024 tokens and repeat 10 times to report average energy and latency values.

\begin{figure}[t]
  \centering
  \includegraphics[width=\linewidth]{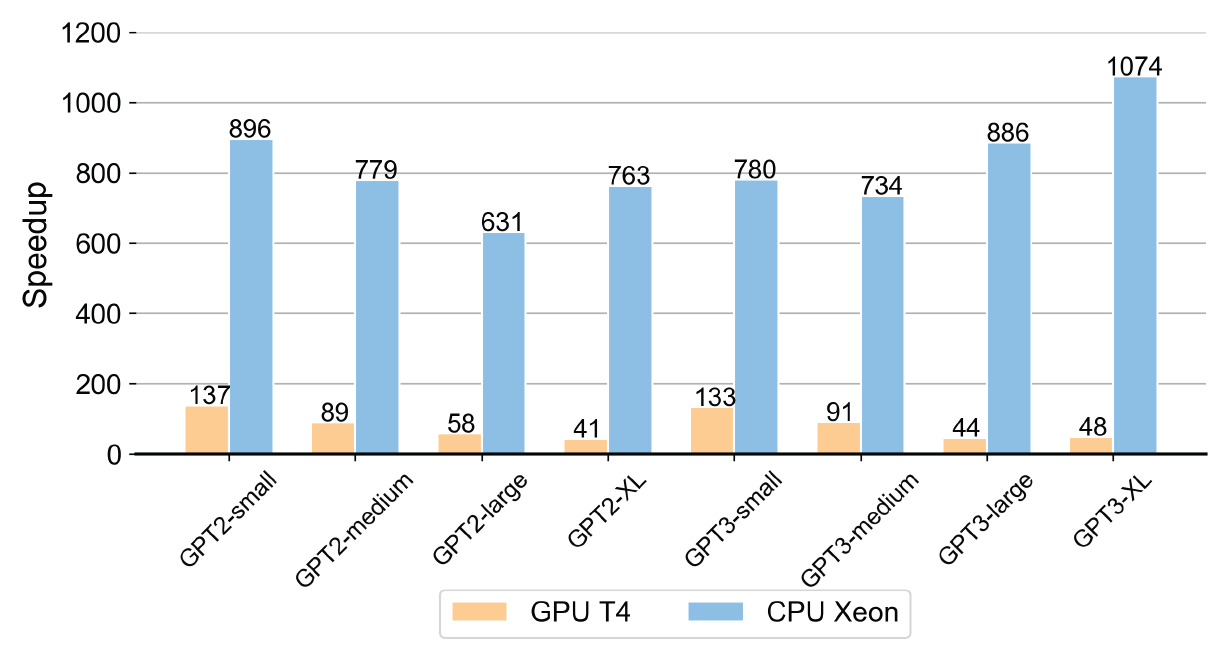}
  \caption{Speedup w.r.t GPU and CPU.}
  \label{fig8}
\end{figure}

\begin{figure}[t]
  \centering
  \includegraphics[width=\linewidth]{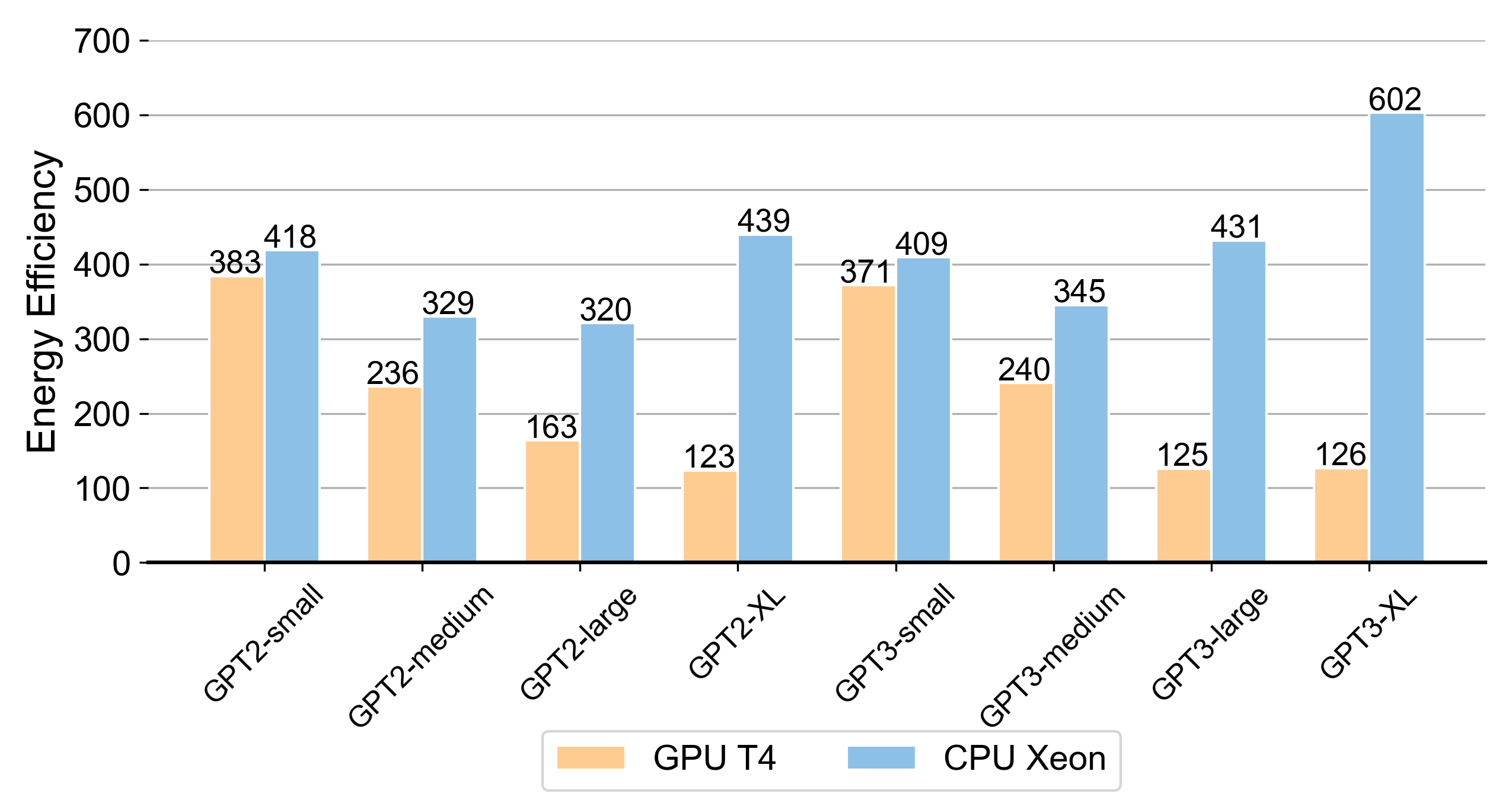}
  \caption{Energy efficiency w.r.t GPU and CPU.}
  \label{fig9}
\end{figure}

\subsection{Overall System Performance}
\textbf{Speedup} 
As shown in Figure \ref{fig8}, the PIM-GPT system achieves remarkable performance improvements, 41$-$137$\times$ speedup over GPU and 639$-$1074$\times$ speedup over CPU for the 8 GPT models. The high speedup originates from three aspects: (1) Memory bottleneck is effectively removed by performing the memory-intensive VMM operations inside PIM channel; (2) The mapping strategy maximizes computation parallelism and data locality; (3) Different workloads are efficiently distributed between PIM and ASIC. In comparison, GPU is not suitable for sequential token generation, since the large memory footprint and low data reuse rate under-utilize the GPU computation resources \cite{hong2022dfx}. 

We also compare the PIM-GPT performance with previously reported Transformer accelerators, as shown in Table \ref{tab:2}. SpAtten \cite{wang2021spatten} accelerates GPT2-medium by 35$\times$ over GPU, and TransPIM \cite{zhou2022transpim} obtains similar speedup of 33$\times$. Both ignore the layer normalization and residual connections in Transformer models. DFX \cite{hong2022dfx} provides 3.2$\times$ latency reduction on average. PIM-GPT achieves state-of-the-art performance with 89$\times$ speedup on average over GPU. We also want to highlight that the PIM-GPT testing result is based on 1024 tokens, which cannot be supported by these prior prototypes. 

\textbf{Energy Efficiency}
Figure \ref{fig9} shows that the PIM-GPT achieves energy reduction of 123$-$383$\times$ and 320$-$602$\times$ over GPU and CPU, respectively. PIM-GPT effectively eliminates the energy consumption of DRAM data transmission by using PIM to locally consume data. Additionally, the mapping method leverages data locality and minimizes the row ACT and PRE operations that are energy consuming. The ASIC only contributes a very small fraction of the total system energy, but provides highly efficient arithmetic computations. 

In comparison, DFX only achieves 3.99$\times$ higher energy efficiency compared to the GPU baseline \cite{hong2022dfx}. TransPIM reports $\sim$250$\times$ energy reduction \cite{zhou2022transpim}. SpAtten reports 382$\times$ over GPU \cite{wang2021spatten}, but only the attention layer is included in energy estimate while others including FFN layers are not considered. Speedup and energy efficiency for the above mentioned accelerators are summarized in Table \ref{tab:2}.

\begin{table}[t]
  \renewcommand*{\arraystretch}{1.4}
  \setlength{\arrayrulewidth}{0.25mm}
  \centering
  \caption{Comparison with Other GPT Accelerators.}
  \label{table:formatting}
  \begin{tabularx}{\linewidth}{|X|X|X|X|X|}
    \hline
     & SpAtten \newline\cite{wang2021spatten} & TransPIM\newline\cite{zhou2022transpim} & DFX\newline\cite{hong2022dfx} & \textbf{PIM-GPT} \\
    \hline
    Memory & HBM & HBM & HBM & GDDR6 \\
    \hline
    End-to-end & \ding{55} & \ding{55} & \ding{51} & \ding{51}\\
    \hline
    Data Type & INT & INT & FP16 & BF16 \\
    \hline
    Largest Model & GPT2-medium & GPT2-medium & GPT2-XL & GPT2/3-XL \\
    \hline
    Speedup & 35$\times$ & 33$\times$ & 3.2$\times$ & 89$\times$ \\
    \hline
    Energy \newline Efficiency & - & $\sim$250$\times$ & 3.99$\times$ & 221$\times$ \\
    \hline
    \end{tabularx}
    \begin{minipage}{\linewidth}
    \vspace{0.1cm}
    \small  Notes: Speedup and energy efficiency are over GPU. PIM-GPT results are based on 1024 token generation.
\end{minipage}
  \label{tab:2}
\end{table}

\subsection{Detailed Performance Analysis} 

\begin{figure}[t]
  \centering
  \includegraphics[width=\linewidth]{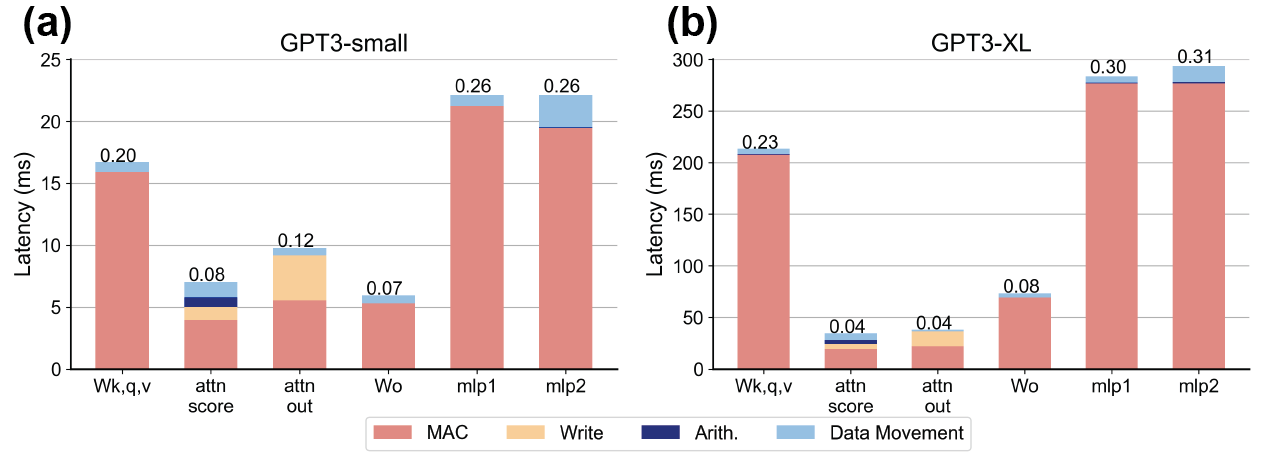}
  \caption{Layer-wise latency breakdown of (a) GPT3-small and (b) GPT3-XL.}
  \label{fig10}
\end{figure}

The layerwise latency breakdown of GPT3-small and GPT3-XL in Figure \ref{fig10} shows that VMM operations dominate the total execution time of PIM-GPT. All other arithmetic computations only account for 1.16\% of total latency in GPT3-XL. For larger Transformer models, the improvement of PIM-GPT over GPU is reduced. This is because larger GPT models allow better utilization of GPU computation resources. As a result, the gain over GPU is reduced when compared with smaller GPT models, although the performance gain ($>$40$\times$) is still significant.  

\begin{figure}[h]
  \centering
  \includegraphics[width=\linewidth]{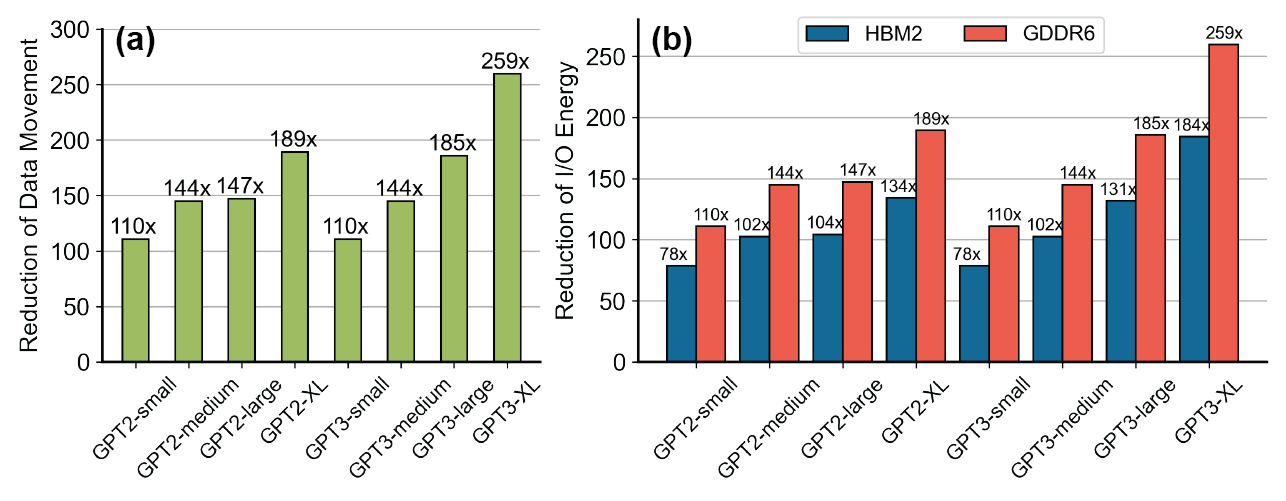}
  \caption{(a) Reduction of data movement. (b) Reduction of DRAM I/O energy consumption.}
  \label{fig11}
\end{figure}

For PIM-GPT, the VMM computation occurs on the same DRAM chip that stores the required weights and Key, Value mactrices. Therefore, a significant amount of data movement can be eliminated. The data movement only happens when VMM results are transmitted to the ASIC for downstream processing or data synchronization. Figure \ref{fig11}(a) shows the data transfer reduction can be 110$-$259$\times$. In PIM-GPT, data movement no longer becomes the bottleneck and consumes a very small proportion of the total latency, as illustrated by the layerwise breakdown in Figure \ref{fig10}. We also evaluated the reduction of energy consumption of DRAM I/O compared to GPU based on HBM2 and GDDR6 memory, which is commonly used in the state-of-the-art GPUs. The HBM2 access energy 3.9 pJ/bit is adopted from \cite{o2017fine}. PIM-GPT achieves 78$-$184$\times$ and 110$-$259$\times$ I/O energy reduction with respect to HBM2 and GDDR6, as shown in Figure \ref{fig11}(b), proving the effectiveness of PIM-GPT in eliminating external matrix data movement.

\begin{figure}[t]
  \centering
  \includegraphics[width=\linewidth]{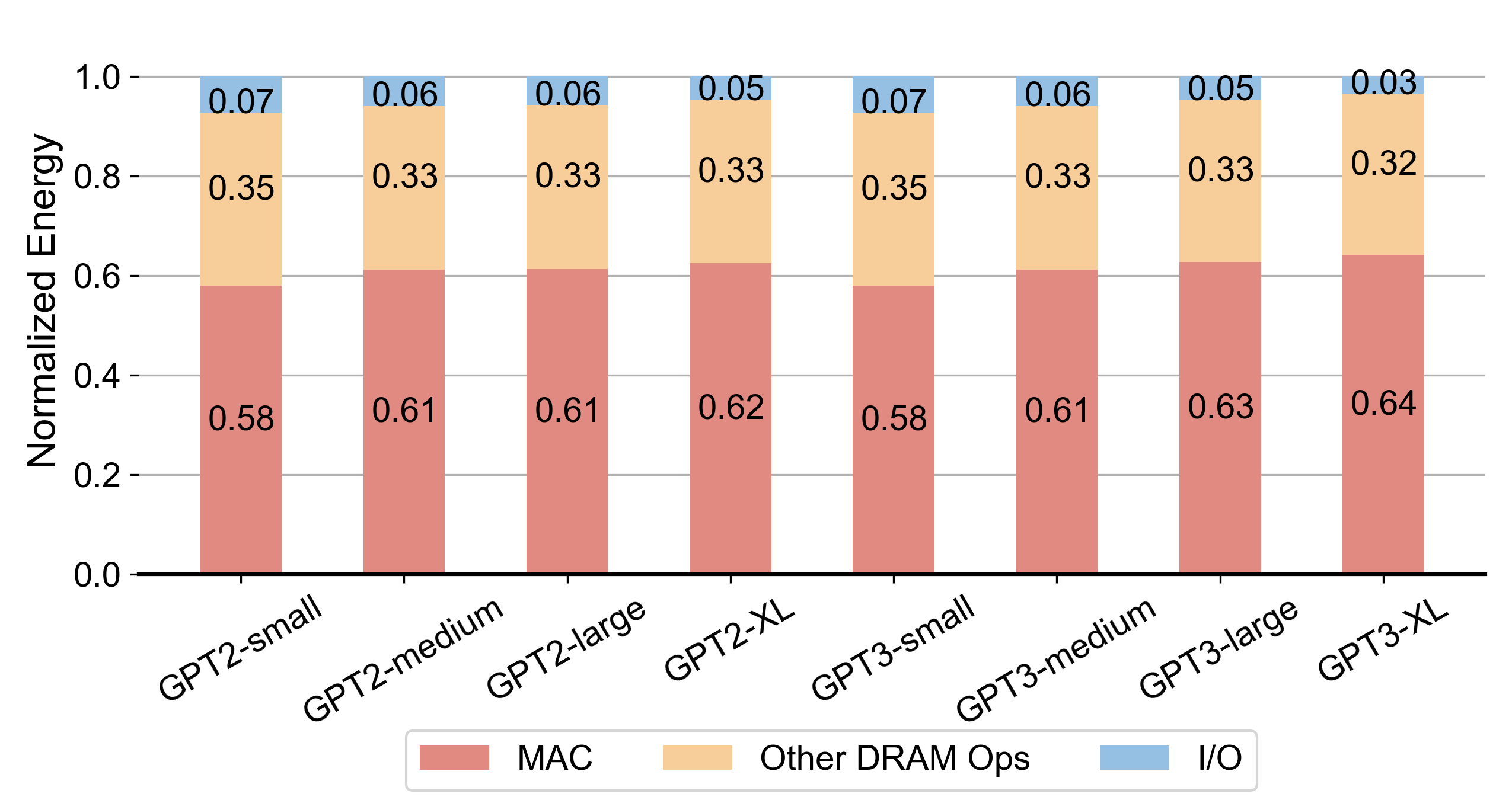}
  \caption{DRAM energy consumption breakdown of all models.}
  \label{fig12}
\end{figure}

Since weight matrices reside in DRAM, only the input and output vectors will be transferred though the I/O. Hence, the memory access complexity can be reduced from $\mathcal{O}(n^2)$ to $\mathcal{O}(n)$. As a result, the energy consumption through DRAM I/O in PIM-GPT is less then 10\% of total DRAM energy consumption in PIM-GPT, as shown in DRAM energy consumption breakdown in Figure \ref{fig12}. MAC operation consumes most of the DRAM energy, since VMM is the core part of the GPT. Other DRAM operations, such as ACT, PRE and REF, along with the standby energy consumption consumes around 33\% of DRAM energy.

\begin{figure}[h]
  \centering
  \includegraphics[width=\linewidth]{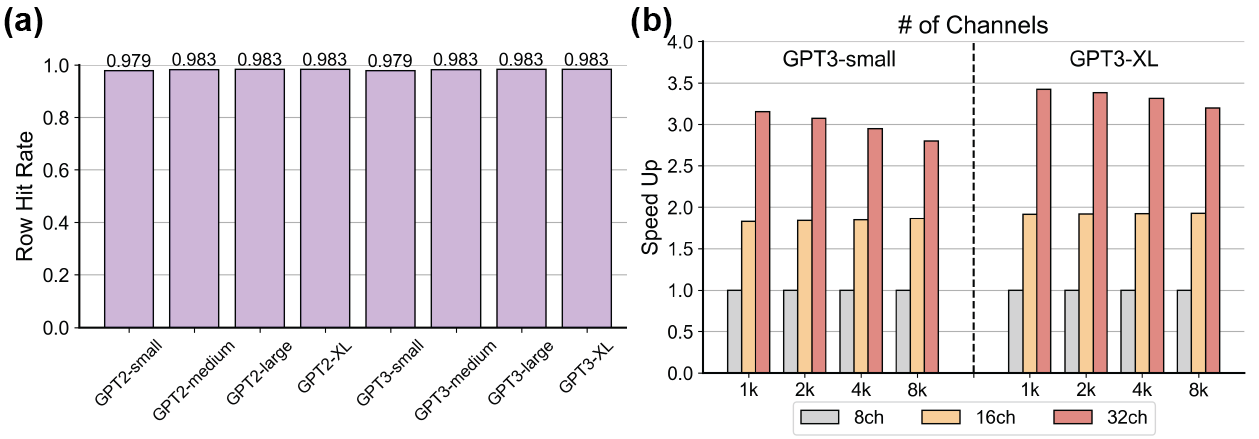}
  \caption{(a) Row hit rate. (b) PIM channel scalability.}
  \label{fig13}
\end{figure}

In PIM-GPT, data locality is optimized during model mapping to reduce the memory access time and enhance the computation throughput, which is evaluated by row hit rate. If a data is access from the row buffer without activating the bank row, such an access is a hit. The row hit rate is calculated by the number of hits over total data access. Figure \ref{fig13}(a) plots the row hit rates, achieving $\sim98\%$ for the 8 GPT models. The system performance is further enhanced by increasing computation parallelism, as shown in Figure \ref{fig13}(b). More memory channels can be attached to the ASIC with relatively minor modifications to the ASIC data port. The improvement scales almost linearly with the number of channels, proving the efficacy of parallelism-aware mapping scheme in PIM-GPT. The speedup slightly reduces for longer sequences, because longer sequences require more arithmetic computation on ASIC. 

\subsection{Sensitivity Study}
\begin{figure}[t]
  \centering
  \includegraphics[width=\linewidth]{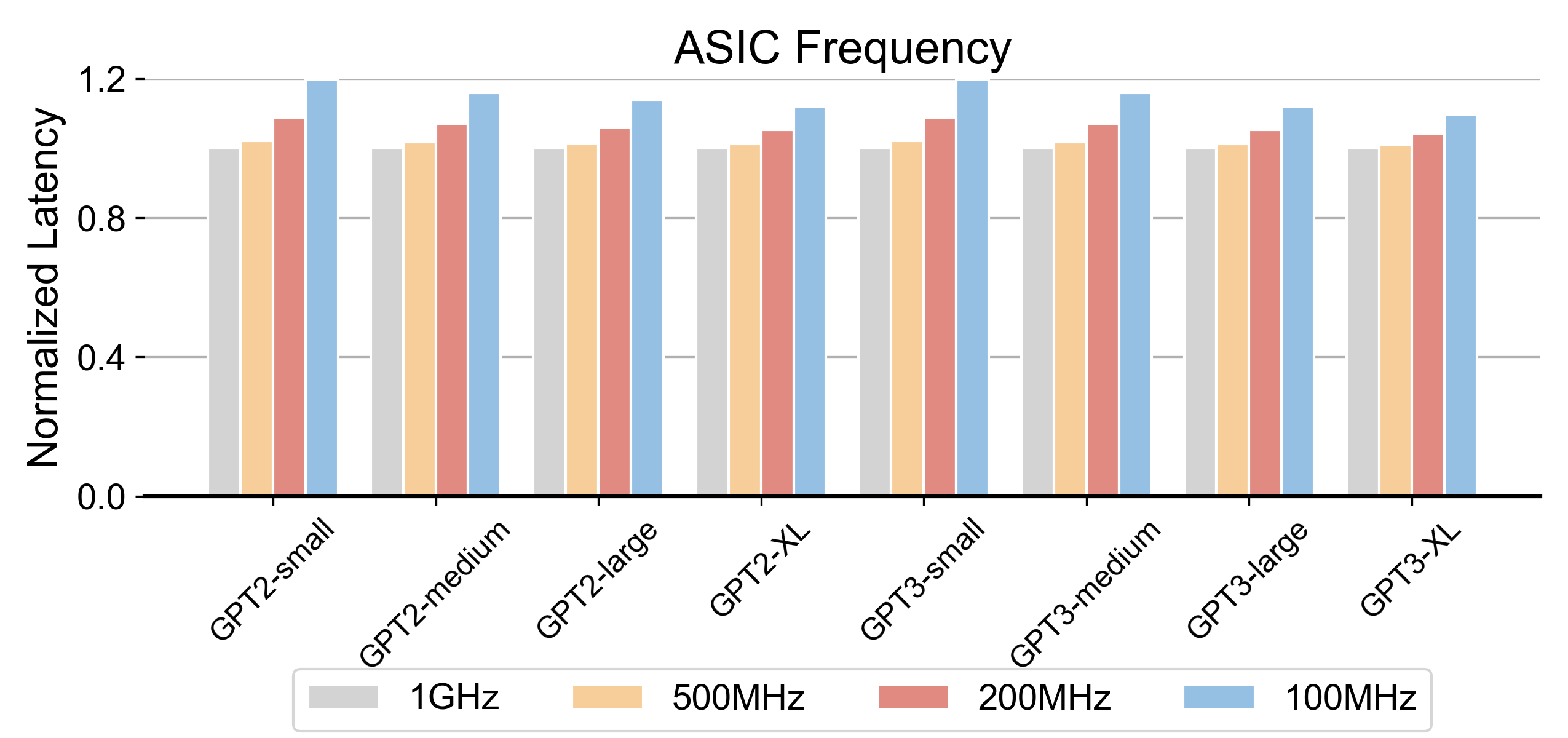}
  \caption{Sensitivity of performance to clock frequency.}
  \label{fig14}
\end{figure}

\textbf{ASIC Frequency} 
The PIM-GPT ASIC is designed with TSMC 28~nm technology at 1~GHz clock frequency. However, frequency scaling is an important technique for power optimization. We conduct a sensitivity study of ASIC frequency by varying the latency setting in our simulator. Figure \ref{fig14} shows the latency at different clock frequencies for the 8 GPT models, where the latency is normalized with respect to the 1~GHz results. Overall there is only a small latency increase when the ASIC frequency scales down from 1~GHz to 200~MHz for all models. Even when further scaling the frequency down to 100~MHz, which is 10 times slower than the baseline, the worst case only incurs a performance slowdown of 20\%. Moreover, larger models are less sensitive to the ASIC frequency scaling, since their operations are more dominated by VMM and the proportion of ASIC arithmetic computation is less than in smaller models. Therefore, the PIM-GPT design is not sensitive to ASIC clock frequency, which justifies its use for where power needs to be optimized by reducing clock frequencies.

\begin{figure}[t]
  \centering
  \includegraphics[width=\linewidth]{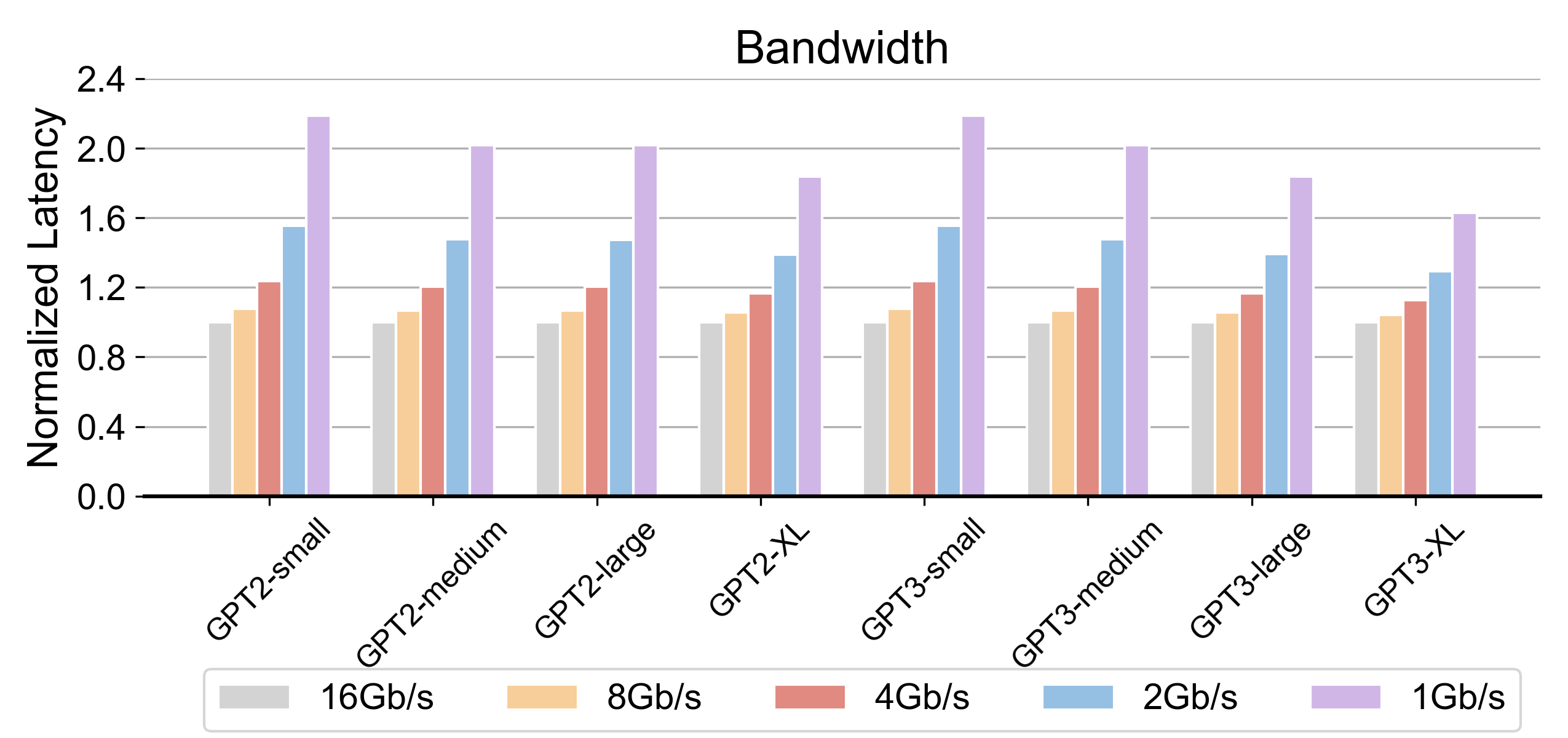}
  \caption{Sensitivity of performance to data transfer rate.}
  \label{fig15}
\end{figure}

\textbf{Data Transmission Rate} 
The memory interface can be a bottleneck for many computation tasks, even for PIM implementations. In \cite{kwon2022system}, SK Hynix reported accelerating fully connected layers in GPT models using GDDR6-PIM system. The system with 4 channels experienced $\sim$3$\times$ slowdown when memory interface bandwidth changes from 16~Gb/s/pin to 2~Gb/s/pin. We test the PIM-GPT's sensitivity to the memory interface by changing the bandwidth configuration in our simulator. Figure \ref{fig15} shows the latency as a function of memory interface bandwidth for the 8 GPT models. When the memory interface bandwidth changes from 16~Gb/s/pin to 2~Gb/s/pin, the end-to-end GPT inference time is increased $\sim$1.5$\times$ on average. That is 2$\times$ better than reported in \cite{kwon2022system}, where only VMM and GELU in GPT are considered. Even when the data transfer rate is decreased to 1~Gb/s, all models are slowed down by $\sim$2$\times$ on average. These results show PIM-GPT is not sensitive to the memory interface bandwidth, since most data are consumed locally on-chip and the data transfer requirements are significantly reduced. 

\section{Related Work}
\textbf{Process-in-Memory}
PIM architectures optimize the resolution of memory-bounded tasks by positioning processing units in close proximity to data, leveraging high internal bandwidth. DRAM-based PIM is one of the most promising technologies to accelerate data-intensive computing tasks. Besides research advances, DRAM vendors including Samsung \cite{kwon202125}\cite{lee2021hardware} and SK Hynix \cite{lee20221ynm}\cite{kwon20221ynm} have recently announced DRAM-based PIM technologies. Samsung's PIM architecture is based on HBM2, which offers 307.2~GBps bandwidth to tackle data intensive tasks. The design integrates PIM dies on a buffer die through TSV. Inside each PIM die, a PU is shared by two banks, operating at 9.6GFLOPS per PU. However, the high costs associated with through-silicon vias (TSV) make this approach less cost-effective. SK Hynix's GDDR6-based PIM prototype, Accelerator-in-Memory (AiM), supports VMM with high throughput of 32GFLOPS per PU. But AiM cannot accelerate end-to-end applications standalone due to the limited functions on DRAM chips. Compared to AiM, PIM-GPT limits PIM to only VMM operations, and performs the other arithmetic and control functions in a separate ASIC with optimized mapping and dataflow designs to allow efficient end-to-end GPT acceleration.  

\textbf{GPT Accelerators}
Recently, several Transformer accelerators have been proposed. Most of them target computation-intensive encoder models \cite{ham20203}\cite{ham2021elsa}\cite{jang2019mnnfast}\cite{zadeh2020gobo} and are not well-optimized for the memory-bounded GPT. Few works accelerate GPT token generation tasks \cite{wang2021spatten}\cite{hong2022dfx}\cite{zhou2022transpim}. DFX \cite{hong2022dfx} partitions the GPT models to multiple FPGAs to leverage parallel computation, but does not reduce the data load/store. SpAtten \cite{wang2021spatten} involves algorithmic optimization such as pruning and quantization to reduce memory overhead. However, these methods inevitably reduce the model inference accuracy. And SpAtten only speeds up the attention mechanism in Transformer. TransPIM \cite{zhou2022transpim} takes advantage of PIM and proposes a token-based dataflow requiring the ring broadcast buffer (2kb) per bank to store temporary data. The design requires subarray-level modification and extensive on-DRAM data movement. The proposed PIM-GPT requires minimal changes to DRAM and optimizes the workload distribution to achieve fast end-to-end GPT acceleration. 

\section{Conclusion}
In this work, we propose a complete, hybrid system, PIM-GPT, to accelerate memory-bounded GPT token generation tasks. PIM-GPT consists of PIM chips and a lightweight ASIC to support end-to-end GPT acceleration. Model mapping and workload distribution are optimized to maximize computation parallelism and data locality. The proposed system achieves 41$-$137$\times$, 631$-$1074$\times$ speedup and 123$-$383$\times$, 320$-$602$\times$ energy efficiency over GPU and CPU on 8 GPT models by eliminating matrix data movement. We highlight that the design only requires light modifications to the DRAM architecture and thus offers a practical and efficient PIM solution.

\section{Acknowledgments}
This work was supported in part by the Semiconductor Research Corporation (SRC) and Defense Advanced Research Projects Agency (DARPA) through the Applications Driving Architectures (ADA) Research Center and in part by the National Science Foundation under Grant CCF-1900675 and ECCS-1915550.

\bibliographystyle{ACM-Reference-Format}
\bibliography{references}


\begin{thebibliography}{44}


\ifx \showCODEN    \undefined \def \showCODEN     #1{\unskip}     \fi
\ifx \showDOI      \undefined \def \showDOI       #1{#1}\fi
\ifx \showISBNx    \undefined \def \showISBNx     #1{\unskip}     \fi
\ifx \showISBNxiii \undefined \def \showISBNxiii  #1{\unskip}     \fi
\ifx \showISSN     \undefined \def \showISSN      #1{\unskip}     \fi
\ifx \showLCCN     \undefined \def \showLCCN      #1{\unskip}     \fi
\ifx \shownote     \undefined \def \shownote      #1{#1}          \fi
\ifx \showarticletitle \undefined \def \showarticletitle #1{#1}   \fi
\ifx \showURL      \undefined \def \showURL       {\relax}        \fi
\providecommand\bibfield[2]{#2}
\providecommand\bibinfo[2]{#2}
\providecommand\natexlab[1]{#1}
\providecommand\showeprint[2][]{arXiv:#2}

\bibitem[Arena(1999)]%
        {invsqrt}
\bibfield{author}{\bibinfo{person}{Quake~III Arena}.} \bibinfo{year}{1999}\natexlab{}.
\newblock \bibinfo{title}{Fast inverse square root}.
\newblock
\newblock
\urldef\tempurl%
\url{https://en.wikipedia.org/wiki/Fast_inverse_square_root}
\showURL{%
Retrieved July 2023 from \tempurl}


\bibitem[Brown et~al\mbox{.}(2020)]%
        {brown2020language}
\bibfield{author}{\bibinfo{person}{Tom Brown}, \bibinfo{person}{Benjamin Mann}, \bibinfo{person}{Nick Ryder}, \bibinfo{person}{Melanie Subbiah}, \bibinfo{person}{Jared~D Kaplan}, \bibinfo{person}{Prafulla Dhariwal}, \bibinfo{person}{Arvind Neelakantan}, \bibinfo{person}{Pranav Shyam}, \bibinfo{person}{Girish Sastry}, \bibinfo{person}{Amanda Askell}, {et~al\mbox{.}}} \bibinfo{year}{2020}\natexlab{}.
\newblock \showarticletitle{Language models are few-shot learners}.
\newblock \bibinfo{journal}{\emph{Advances in neural information processing systems}}  \bibinfo{volume}{33} (\bibinfo{year}{2020}), \bibinfo{pages}{1877--1901}.
\newblock


\bibitem[Chang et~al\mbox{.}(2020)]%
        {chang2020taming}
\bibfield{author}{\bibinfo{person}{Wei-Cheng Chang}, \bibinfo{person}{Hsiang-Fu Yu}, \bibinfo{person}{Kai Zhong}, \bibinfo{person}{Yiming Yang}, {and} \bibinfo{person}{Inderjit~S Dhillon}.} \bibinfo{year}{2020}\natexlab{}.
\newblock \showarticletitle{Taming pretrained transformers for extreme multi-label text classification}. In \bibinfo{booktitle}{\emph{Proceedings of the 26th ACM SIGKDD international conference on knowledge discovery \& data mining}}. \bibinfo{pages}{3163--3171}.
\newblock


\bibitem[Dass et~al\mbox{.}(2023)]%
        {dass2023vitality}
\bibfield{author}{\bibinfo{person}{Jyotikrishna Dass}, \bibinfo{person}{Shang Wu}, \bibinfo{person}{Huihong Shi}, \bibinfo{person}{Chaojian Li}, \bibinfo{person}{Zhifan Ye}, \bibinfo{person}{Zhongfeng Wang}, {and} \bibinfo{person}{Yingyan Lin}.} \bibinfo{year}{2023}\natexlab{}.
\newblock \showarticletitle{Vitality: Unifying low-rank and sparse approximation for vision transformer acceleration with a linear taylor attention}. In \bibinfo{booktitle}{\emph{2023 IEEE International Symposium on High-Performance Computer Architecture (HPCA)}}. IEEE, \bibinfo{pages}{415--428}.
\newblock


\bibitem[Devaux(2019)]%
        {devaux2019true}
\bibfield{author}{\bibinfo{person}{Fabrice Devaux}.} \bibinfo{year}{2019}\natexlab{}.
\newblock \showarticletitle{The true processing in memory accelerator}. In \bibinfo{booktitle}{\emph{2019 IEEE Hot Chips 31 Symposium (HCS)}}. IEEE Computer Society, \bibinfo{pages}{1--24}.
\newblock


\bibitem[Devlin et~al\mbox{.}(2018)]%
        {devlin2018bert}
\bibfield{author}{\bibinfo{person}{Jacob Devlin}, \bibinfo{person}{Ming-Wei Chang}, \bibinfo{person}{Kenton Lee}, {and} \bibinfo{person}{Kristina Toutanova}.} \bibinfo{year}{2018}\natexlab{}.
\newblock \showarticletitle{Bert: Pre-training of deep bidirectional transformers for language understanding}.
\newblock \bibinfo{journal}{\emph{arXiv preprint arXiv:1810.04805}} (\bibinfo{year}{2018}).
\newblock


\bibitem[Fan et~al\mbox{.}(2023)]%
        {fan2023taskfusion}
\bibfield{author}{\bibinfo{person}{Zichen Fan}, \bibinfo{person}{Qirui Zhang}, \bibinfo{person}{Pierre Abillama}, \bibinfo{person}{Sara Shoouri}, \bibinfo{person}{Changwoo Lee}, \bibinfo{person}{David Blaauw}, \bibinfo{person}{Hun-Seok Kim}, {and} \bibinfo{person}{Dennis Sylvester}.} \bibinfo{year}{2023}\natexlab{}.
\newblock \showarticletitle{TaskFusion: An Efficient Transfer Learning Architecture with Dual Delta Sparsity for Multi-Task Natural Language Processing}. In \bibinfo{booktitle}{\emph{Proceedings of the 50th Annual International Symposium on Computer Architecture}}. \bibinfo{pages}{1--14}.
\newblock


\bibitem[Gao et~al\mbox{.}(2019)]%
        {gao2019computedram}
\bibfield{author}{\bibinfo{person}{Fei Gao}, \bibinfo{person}{Georgios Tziantzioulis}, {and} \bibinfo{person}{David Wentzlaff}.} \bibinfo{year}{2019}\natexlab{}.
\newblock \showarticletitle{Computedram: In-memory compute using off-the-shelf drams}. In \bibinfo{booktitle}{\emph{Proceedings of the 52nd annual IEEE/ACM international symposium on microarchitecture}}. \bibinfo{pages}{100--113}.
\newblock


\bibitem[Garg and Ramakrishnan(2020)]%
        {garg2020bae}
\bibfield{author}{\bibinfo{person}{Siddhant Garg} {and} \bibinfo{person}{Goutham Ramakrishnan}.} \bibinfo{year}{2020}\natexlab{}.
\newblock \showarticletitle{Bae: Bert-based adversarial examples for text classification}.
\newblock \bibinfo{journal}{\emph{arXiv preprint arXiv:2004.01970}} (\bibinfo{year}{2020}).
\newblock


\bibitem[Ghose et~al\mbox{.}(2019)]%
        {ghose2019processing}
\bibfield{author}{\bibinfo{person}{Saugata Ghose}, \bibinfo{person}{Amirali Boroumand}, \bibinfo{person}{Jeremie~S Kim}, \bibinfo{person}{Juan G{\'o}mez-Luna}, {and} \bibinfo{person}{Onur Mutlu}.} \bibinfo{year}{2019}\natexlab{}.
\newblock \showarticletitle{Processing-in-memory: A workload-driven perspective}.
\newblock \bibinfo{journal}{\emph{IBM Journal of Research and Development}} \bibinfo{volume}{63}, \bibinfo{number}{6} (\bibinfo{year}{2019}), \bibinfo{pages}{3--1}.
\newblock


\bibitem[Ghose et~al\mbox{.}(2018)]%
        {ghose2018your}
\bibfield{author}{\bibinfo{person}{Saugata Ghose}, \bibinfo{person}{Abdullah~Giray Yaglik{\c{c}}i}, \bibinfo{person}{Raghav Gupta}, \bibinfo{person}{Donghyuk Lee}, \bibinfo{person}{Kais Kudrolli}, \bibinfo{person}{William~X Liu}, \bibinfo{person}{Hasan Hassan}, \bibinfo{person}{Kevin~K Chang}, \bibinfo{person}{Niladrish Chatterjee}, \bibinfo{person}{Aditya Agrawal}, {et~al\mbox{.}}} \bibinfo{year}{2018}\natexlab{}.
\newblock \showarticletitle{What your DRAM power models are not telling you: Lessons from a detailed experimental study}.
\newblock \bibinfo{journal}{\emph{Proceedings of the ACM on Measurement and Analysis of Computing Systems}} \bibinfo{volume}{2}, \bibinfo{number}{3} (\bibinfo{year}{2018}), \bibinfo{pages}{1--41}.
\newblock


\bibitem[G{\'o}mez-Luna et~al\mbox{.}(2022)]%
        {gomez2022benchmarking}
\bibfield{author}{\bibinfo{person}{Juan G{\'o}mez-Luna}, \bibinfo{person}{Izzat El~Hajj}, \bibinfo{person}{Ivan Fernandez}, \bibinfo{person}{Christina Giannoula}, \bibinfo{person}{Geraldo~F Oliveira}, {and} \bibinfo{person}{Onur Mutlu}.} \bibinfo{year}{2022}\natexlab{}.
\newblock \showarticletitle{Benchmarking a new paradigm: Experimental analysis and characterization of a real processing-in-memory system}.
\newblock \bibinfo{journal}{\emph{IEEE Access}}  \bibinfo{volume}{10} (\bibinfo{year}{2022}), \bibinfo{pages}{52565--52608}.
\newblock


\bibitem[Ham et~al\mbox{.}(2020)]%
        {ham20203}
\bibfield{author}{\bibinfo{person}{Tae~Jun Ham}, \bibinfo{person}{Sung~Jun Jung}, \bibinfo{person}{Seonghak Kim}, \bibinfo{person}{Young~H Oh}, \bibinfo{person}{Yeonhong Park}, \bibinfo{person}{Yoonho Song}, \bibinfo{person}{Jung-Hun Park}, \bibinfo{person}{Sanghee Lee}, \bibinfo{person}{Kyoung Park}, \bibinfo{person}{Jae~W Lee}, {et~al\mbox{.}}} \bibinfo{year}{2020}\natexlab{}.
\newblock \showarticletitle{A\^{} 3: Accelerating attention mechanisms in neural networks with approximation}. In \bibinfo{booktitle}{\emph{2020 IEEE International Symposium on High Performance Computer Architecture (HPCA)}}. IEEE, \bibinfo{pages}{328--341}.
\newblock


\bibitem[Ham et~al\mbox{.}(2021)]%
        {ham2021elsa}
\bibfield{author}{\bibinfo{person}{Tae~Jun Ham}, \bibinfo{person}{Yejin Lee}, \bibinfo{person}{Seong~Hoon Seo}, \bibinfo{person}{Soosung Kim}, \bibinfo{person}{Hyunji Choi}, \bibinfo{person}{Sung~Jun Jung}, {and} \bibinfo{person}{Jae~W Lee}.} \bibinfo{year}{2021}\natexlab{}.
\newblock \showarticletitle{ELSA: Hardware-software co-design for efficient, lightweight self-attention mechanism in neural networks}. In \bibinfo{booktitle}{\emph{2021 ACM/IEEE 48th Annual International Symposium on Computer Architecture (ISCA)}}. IEEE, \bibinfo{pages}{692--705}.
\newblock


\bibitem[He et~al\mbox{.}(2016)]%
        {he2016deep}
\bibfield{author}{\bibinfo{person}{Kaiming He}, \bibinfo{person}{Xiangyu Zhang}, \bibinfo{person}{Shaoqing Ren}, {and} \bibinfo{person}{Jian Sun}.} \bibinfo{year}{2016}\natexlab{}.
\newblock \showarticletitle{Deep residual learning for image recognition}. In \bibinfo{booktitle}{\emph{Proceedings of the IEEE conference on computer vision and pattern recognition}}. \bibinfo{pages}{770--778}.
\newblock


\bibitem[He et~al\mbox{.}(2020)]%
        {he2020newton}
\bibfield{author}{\bibinfo{person}{Mingxuan He}, \bibinfo{person}{Choungki Song}, \bibinfo{person}{Ilkon Kim}, \bibinfo{person}{Chunseok Jeong}, \bibinfo{person}{Seho Kim}, \bibinfo{person}{Il Park}, \bibinfo{person}{Mithuna Thottethodi}, {and} \bibinfo{person}{TN Vijaykumar}.} \bibinfo{year}{2020}\natexlab{}.
\newblock \showarticletitle{Newton: A DRAM-maker’s accelerator-in-memory (AiM) architecture for machine learning}. In \bibinfo{booktitle}{\emph{2020 53rd Annual IEEE/ACM International Symposium on Microarchitecture (MICRO)}}. IEEE, \bibinfo{pages}{372--385}.
\newblock


\bibitem[Hendrycks and Gimpel(2016)]%
        {hendrycks2016gaussian}
\bibfield{author}{\bibinfo{person}{Dan Hendrycks} {and} \bibinfo{person}{Kevin Gimpel}.} \bibinfo{year}{2016}\natexlab{}.
\newblock \showarticletitle{Gaussian error linear units (gelus)}.
\newblock \bibinfo{journal}{\emph{arXiv preprint arXiv:1606.08415}} (\bibinfo{year}{2016}).
\newblock


\bibitem[Hong et~al\mbox{.}(2022)]%
        {hong2022dfx}
\bibfield{author}{\bibinfo{person}{Seongmin Hong}, \bibinfo{person}{Seungjae Moon}, \bibinfo{person}{Junsoo Kim}, \bibinfo{person}{Sungjae Lee}, \bibinfo{person}{Minsub Kim}, \bibinfo{person}{Dongsoo Lee}, {and} \bibinfo{person}{Joo-Young Kim}.} \bibinfo{year}{2022}\natexlab{}.
\newblock \showarticletitle{DFX: A Low-latency Multi-FPGA Appliance for Accelerating Transformer-based Text Generation}. In \bibinfo{booktitle}{\emph{2022 55th IEEE/ACM International Symposium on Microarchitecture (MICRO)}}. IEEE, \bibinfo{pages}{616--630}.
\newblock


\bibitem[Jang et~al\mbox{.}(2019)]%
        {jang2019mnnfast}
\bibfield{author}{\bibinfo{person}{Hanhwi Jang}, \bibinfo{person}{Joonsung Kim}, \bibinfo{person}{Jae-Eon Jo}, \bibinfo{person}{Jaewon Lee}, {and} \bibinfo{person}{Jangwoo Kim}.} \bibinfo{year}{2019}\natexlab{}.
\newblock \showarticletitle{Mnnfast: A fast and scalable system architecture for memory-augmented neural networks}. In \bibinfo{booktitle}{\emph{Proceedings of the 46th International Symposium on Computer Architecture}}. \bibinfo{pages}{250--263}.
\newblock


\bibitem[Jeong and Jung(2022)]%
        {jeong2022mac}
\bibfield{author}{\bibinfo{person}{Minki Jeong} {and} \bibinfo{person}{Wanyeong Jung}.} \bibinfo{year}{2022}\natexlab{}.
\newblock \showarticletitle{MAC-DO: Charge Based Multi-Bit Analog In-Memory Accelerator Compatible with DRAM Using Output Stationary Mapping}.
\newblock \bibinfo{journal}{\emph{arXiv preprint arXiv:2207.07862}} (\bibinfo{year}{2022}).
\newblock


\bibitem[Keckler et~al\mbox{.}(2011)]%
        {keckler2011gpus}
\bibfield{author}{\bibinfo{person}{Stephen~W Keckler}, \bibinfo{person}{William~J Dally}, \bibinfo{person}{Brucek Khailany}, \bibinfo{person}{Michael Garland}, {and} \bibinfo{person}{David Glasco}.} \bibinfo{year}{2011}\natexlab{}.
\newblock \showarticletitle{GPUs and the future of parallel computing}.
\newblock \bibinfo{journal}{\emph{IEEE micro}} \bibinfo{volume}{31}, \bibinfo{number}{5} (\bibinfo{year}{2011}), \bibinfo{pages}{7--17}.
\newblock


\bibitem[Kim et~al\mbox{.}(2015)]%
        {kim2015ramulator}
\bibfield{author}{\bibinfo{person}{Yoongu Kim}, \bibinfo{person}{Weikun Yang}, {and} \bibinfo{person}{Onur Mutlu}.} \bibinfo{year}{2015}\natexlab{}.
\newblock \showarticletitle{Ramulator: A fast and extensible DRAM simulator}.
\newblock \bibinfo{journal}{\emph{IEEE Computer architecture letters}} \bibinfo{volume}{15}, \bibinfo{number}{1} (\bibinfo{year}{2015}), \bibinfo{pages}{45--49}.
\newblock


\bibitem[Kwon et~al\mbox{.}(2022a)]%
        {kwon20221ynm}
\bibfield{author}{\bibinfo{person}{Daehan Kwon}, \bibinfo{person}{Seongju Lee}, \bibinfo{person}{Kyuyoung Kim}, \bibinfo{person}{Sanghoon Oh}, \bibinfo{person}{Joonhong Park}, \bibinfo{person}{Gi-Moon Hong}, \bibinfo{person}{Dongyoon Ka}, \bibinfo{person}{Kyudong Hwang}, \bibinfo{person}{Jeongje Park}, \bibinfo{person}{Kyeongpil Kang}, {et~al\mbox{.}}} \bibinfo{year}{2022}\natexlab{a}.
\newblock \showarticletitle{A 1ynm 1.25 V 8Gb 16Gb/s/Pin GDDR6-Based Accelerator-in-Memory Supporting 1TFLOPS MAC Operation and Various Activation Functions for Deep Learning Application}.
\newblock \bibinfo{journal}{\emph{IEEE Journal of Solid-State Circuits}} \bibinfo{volume}{58}, \bibinfo{number}{1} (\bibinfo{year}{2022}), \bibinfo{pages}{291--302}.
\newblock


\bibitem[Kwon et~al\mbox{.}(2022b)]%
        {kwon2022system}
\bibfield{author}{\bibinfo{person}{Yongkee Kwon}, \bibinfo{person}{Kornijcuk Vladimir}, \bibinfo{person}{Nahsung Kim}, \bibinfo{person}{Woojae Shin}, \bibinfo{person}{Jongsoon Won}, \bibinfo{person}{Minkyu Lee}, \bibinfo{person}{Hyunha Joo}, \bibinfo{person}{Haerang Choi}, \bibinfo{person}{Guhyun Kim}, \bibinfo{person}{Byeongju An}, {et~al\mbox{.}}} \bibinfo{year}{2022}\natexlab{b}.
\newblock \showarticletitle{System architecture and software stack for GDDR6-AiM}. In \bibinfo{booktitle}{\emph{2022 IEEE Hot Chips 34 Symposium (HCS)}}. IEEE, \bibinfo{pages}{1--25}.
\newblock


\bibitem[Kwon et~al\mbox{.}(2021)]%
        {kwon202125}
\bibfield{author}{\bibinfo{person}{Young-Cheon Kwon}, \bibinfo{person}{Suk~Han Lee}, \bibinfo{person}{Jaehoon Lee}, \bibinfo{person}{Sang-Hyuk Kwon}, \bibinfo{person}{Je~Min Ryu}, \bibinfo{person}{Jong-Pil Son}, \bibinfo{person}{O Seongil}, \bibinfo{person}{Hak-Soo Yu}, \bibinfo{person}{Haesuk Lee}, \bibinfo{person}{Soo~Young Kim}, {et~al\mbox{.}}} \bibinfo{year}{2021}\natexlab{}.
\newblock \showarticletitle{25.4 a 20nm 6gb function-in-memory dram, based on hbm2 with a 1.2 tflops programmable computing unit using bank-level parallelism, for machine learning applications}. In \bibinfo{booktitle}{\emph{2021 IEEE International Solid-State Circuits Conference (ISSCC)}}, Vol.~\bibinfo{volume}{64}. IEEE, \bibinfo{pages}{350--352}.
\newblock


\bibitem[Lee et~al\mbox{.}(2021)]%
        {lee2021hardware}
\bibfield{author}{\bibinfo{person}{Sukhan Lee}, \bibinfo{person}{Shin-haeng Kang}, \bibinfo{person}{Jaehoon Lee}, \bibinfo{person}{Hyeonsu Kim}, \bibinfo{person}{Eojin Lee}, \bibinfo{person}{Seungwoo Seo}, \bibinfo{person}{Hosang Yoon}, \bibinfo{person}{Seungwon Lee}, \bibinfo{person}{Kyounghwan Lim}, \bibinfo{person}{Hyunsung Shin}, {et~al\mbox{.}}} \bibinfo{year}{2021}\natexlab{}.
\newblock \showarticletitle{Hardware architecture and software stack for PIM based on commercial DRAM technology: Industrial product}. In \bibinfo{booktitle}{\emph{2021 ACM/IEEE 48th Annual International Symposium on Computer Architecture (ISCA)}}. IEEE, \bibinfo{pages}{43--56}.
\newblock


\bibitem[Lee et~al\mbox{.}(2022)]%
        {lee20221ynm}
\bibfield{author}{\bibinfo{person}{Seongju Lee}, \bibinfo{person}{Kyuyoung Kim}, \bibinfo{person}{Sanghoon Oh}, \bibinfo{person}{Joonhong Park}, \bibinfo{person}{Gimoon Hong}, \bibinfo{person}{Dongyoon Ka}, \bibinfo{person}{Kyudong Hwang}, \bibinfo{person}{Jeongje Park}, \bibinfo{person}{Kyeongpil Kang}, \bibinfo{person}{Jungyeon Kim}, {et~al\mbox{.}}} \bibinfo{year}{2022}\natexlab{}.
\newblock \showarticletitle{A 1ynm 1.25 v 8gb, 16gb/s/pin gddr6-based accelerator-in-memory supporting 1tflops mac operation and various activation functions for deep-learning applications}. In \bibinfo{booktitle}{\emph{2022 IEEE International Solid-State Circuits Conference (ISSCC)}}, Vol.~\bibinfo{volume}{65}. IEEE, \bibinfo{pages}{1--3}.
\newblock


\bibitem[Li et~al\mbox{.}(2017)]%
        {li2017drisa}
\bibfield{author}{\bibinfo{person}{Shuangchen Li}, \bibinfo{person}{Dimin Niu}, \bibinfo{person}{Krishna~T Malladi}, \bibinfo{person}{Hongzhong Zheng}, \bibinfo{person}{Bob Brennan}, {and} \bibinfo{person}{Yuan Xie}.} \bibinfo{year}{2017}\natexlab{}.
\newblock \showarticletitle{Drisa: A dram-based reconfigurable in-situ accelerator}. In \bibinfo{booktitle}{\emph{Proceedings of the 50th Annual IEEE/ACM International Symposium on Microarchitecture}}. \bibinfo{pages}{288--301}.
\newblock


\bibitem[Micron(2017)]%
        {gddr6}
\bibfield{author}{\bibinfo{person}{Micron}.} \bibinfo{year}{2017}\natexlab{}.
\newblock \bibinfo{title}{TN-ED-03: GDDR6: The Next-Generation Graphics DRAM}.
\newblock
\newblock
\urldef\tempurl%
\url{https://www.micron.com/-/media/client/global/documents/products/technical-note/dram/tned03_gddr6.pdf}
\showURL{%
Retrieved Nov 2023 from \tempurl}


\bibitem[Micron(2021)]%
        {micron}
\bibfield{author}{\bibinfo{person}{Micron}.} \bibinfo{year}{2021}\natexlab{}.
\newblock \bibinfo{title}{16Gb DDR5 SDRAM Addendum}.
\newblock
\newblock
\urldef\tempurl%
\url{https://media-www.micron.com/-/media/client/global/documents/products/data-sheet/dram/ddr5/16gb_ddr5_sdram_diereva.pdf}
\showURL{%
Retrieved May 2023 from \tempurl}


\bibitem[Mutlu(2023)]%
        {mutlu2023memory}
\bibfield{author}{\bibinfo{person}{Onur Mutlu}.} \bibinfo{year}{2023}\natexlab{}.
\newblock \showarticletitle{Memory-Centric Computing}.
\newblock \bibinfo{journal}{\emph{arXiv preprint arXiv:2305.20000}} (\bibinfo{year}{2023}).
\newblock


\bibitem[O'Connor et~al\mbox{.}(2017)]%
        {o2017fine}
\bibfield{author}{\bibinfo{person}{Mike O'Connor}, \bibinfo{person}{Niladrish Chatterjee}, \bibinfo{person}{Donghyuk Lee}, \bibinfo{person}{John Wilson}, \bibinfo{person}{Aditya Agrawal}, \bibinfo{person}{Stephen~W Keckler}, {and} \bibinfo{person}{William~J Dally}.} \bibinfo{year}{2017}\natexlab{}.
\newblock \showarticletitle{Fine-grained DRAM: Energy-efficient DRAM for extreme bandwidth systems}. In \bibinfo{booktitle}{\emph{Proceedings of the 50th Annual IEEE/ACM International Symposium on Microarchitecture}}. \bibinfo{pages}{41--54}.
\newblock


\bibitem[OpenAI(2023)]%
        {openai2023gpt}
\bibfield{author}{\bibinfo{person}{R OpenAI}.} \bibinfo{year}{2023}\natexlab{}.
\newblock \showarticletitle{GPT-4 technical report}.
\newblock \bibinfo{journal}{\emph{arXiv}} (\bibinfo{year}{2023}), \bibinfo{pages}{2303--08774}.
\newblock


\bibitem[Radford et~al\mbox{.}(2019)]%
        {radford2019language}
\bibfield{author}{\bibinfo{person}{Alec Radford}, \bibinfo{person}{Jeffrey Wu}, \bibinfo{person}{Rewon Child}, \bibinfo{person}{David Luan}, \bibinfo{person}{Dario Amodei}, \bibinfo{person}{Ilya Sutskever}, {et~al\mbox{.}}} \bibinfo{year}{2019}\natexlab{}.
\newblock \showarticletitle{Language models are unsupervised multitask learners}.
\newblock \bibinfo{journal}{\emph{OpenAI blog}} \bibinfo{volume}{1}, \bibinfo{number}{8} (\bibinfo{year}{2019}), \bibinfo{pages}{9}.
\newblock


\bibitem[Shin et~al\mbox{.}(2018)]%
        {shin2018mcdram}
\bibfield{author}{\bibinfo{person}{Hyunsung Shin}, \bibinfo{person}{Dongyoung Kim}, \bibinfo{person}{Eunhyeok Park}, \bibinfo{person}{Sungho Park}, \bibinfo{person}{Yongsik Park}, {and} \bibinfo{person}{Sungjoo Yoo}.} \bibinfo{year}{2018}\natexlab{}.
\newblock \showarticletitle{McDRAM: Low latency and energy-efficient matrix computations in DRAM}.
\newblock \bibinfo{journal}{\emph{IEEE Transactions on Computer-Aided Design of Integrated Circuits and Systems}} \bibinfo{volume}{37}, \bibinfo{number}{11} (\bibinfo{year}{2018}), \bibinfo{pages}{2613--2622}.
\newblock


\bibitem[Sun et~al\mbox{.}(2019)]%
        {sun2019fine}
\bibfield{author}{\bibinfo{person}{Chi Sun}, \bibinfo{person}{Xipeng Qiu}, \bibinfo{person}{Yige Xu}, {and} \bibinfo{person}{Xuanjing Huang}.} \bibinfo{year}{2019}\natexlab{}.
\newblock \showarticletitle{How to fine-tune bert for text classification?}. In \bibinfo{booktitle}{\emph{Chinese Computational Linguistics: 18th China National Conference, CCL 2019, Kunming, China, October 18--20, 2019, Proceedings 18}}. Springer, \bibinfo{pages}{194--206}.
\newblock


\bibitem[Vaswani et~al\mbox{.}(2017)]%
        {vaswani2017attention}
\bibfield{author}{\bibinfo{person}{Ashish Vaswani}, \bibinfo{person}{Noam Shazeer}, \bibinfo{person}{Niki Parmar}, \bibinfo{person}{Jakob Uszkoreit}, \bibinfo{person}{Llion Jones}, \bibinfo{person}{Aidan~N Gomez}, \bibinfo{person}{{\L}ukasz Kaiser}, {and} \bibinfo{person}{Illia Polosukhin}.} \bibinfo{year}{2017}\natexlab{}.
\newblock \showarticletitle{Attention is all you need}.
\newblock \bibinfo{journal}{\emph{Advances in neural information processing systems}}  \bibinfo{volume}{30} (\bibinfo{year}{2017}).
\newblock


\bibitem[Wang et~al\mbox{.}(2023)]%
        {wang2023cta}
\bibfield{author}{\bibinfo{person}{Haoran Wang}, \bibinfo{person}{Haobo Xu}, \bibinfo{person}{Ying Wang}, {and} \bibinfo{person}{Yinhe Han}.} \bibinfo{year}{2023}\natexlab{}.
\newblock \showarticletitle{CTA: Hardware-Software Co-design for Compressed Token Attention Mechanism}. In \bibinfo{booktitle}{\emph{2023 IEEE International Symposium on High-Performance Computer Architecture (HPCA)}}. IEEE, \bibinfo{pages}{429--441}.
\newblock


\bibitem[Wang et~al\mbox{.}(2021)]%
        {wang2021spatten}
\bibfield{author}{\bibinfo{person}{Hanrui Wang}, \bibinfo{person}{Zhekai Zhang}, {and} \bibinfo{person}{Song Han}.} \bibinfo{year}{2021}\natexlab{}.
\newblock \showarticletitle{Spatten: Efficient sparse attention architecture with cascade token and head pruning}. In \bibinfo{booktitle}{\emph{2021 IEEE International Symposium on High-Performance Computer Architecture (HPCA)}}. IEEE, \bibinfo{pages}{97--110}.
\newblock


\bibitem[Wang et~al\mbox{.}(2019)]%
        {wang2019learning}
\bibfield{author}{\bibinfo{person}{Qiang Wang}, \bibinfo{person}{Bei Li}, \bibinfo{person}{Tong Xiao}, \bibinfo{person}{Jingbo Zhu}, \bibinfo{person}{Changliang Li}, \bibinfo{person}{Derek~F Wong}, {and} \bibinfo{person}{Lidia~S Chao}.} \bibinfo{year}{2019}\natexlab{}.
\newblock \showarticletitle{Learning deep transformer models for machine translation}.
\newblock \bibinfo{journal}{\emph{arXiv preprint arXiv:1906.01787}} (\bibinfo{year}{2019}).
\newblock


\bibitem[Yao and Wan(2020)]%
        {yao2020multimodal}
\bibfield{author}{\bibinfo{person}{Shaowei Yao} {and} \bibinfo{person}{Xiaojun Wan}.} \bibinfo{year}{2020}\natexlab{}.
\newblock \showarticletitle{Multimodal transformer for multimodal machine translation}. In \bibinfo{booktitle}{\emph{Proceedings of the 58th annual meeting of the association for computational linguistics}}. \bibinfo{pages}{4346--4350}.
\newblock


\bibitem[You et~al\mbox{.}(2023)]%
        {you2023vitcod}
\bibfield{author}{\bibinfo{person}{Haoran You}, \bibinfo{person}{Zhanyi Sun}, \bibinfo{person}{Huihong Shi}, \bibinfo{person}{Zhongzhi Yu}, \bibinfo{person}{Yang Zhao}, \bibinfo{person}{Yongan Zhang}, \bibinfo{person}{Chaojian Li}, \bibinfo{person}{Baopu Li}, {and} \bibinfo{person}{Yingyan Lin}.} \bibinfo{year}{2023}\natexlab{}.
\newblock \showarticletitle{Vitcod: Vision transformer acceleration via dedicated algorithm and accelerator co-design}. In \bibinfo{booktitle}{\emph{2023 IEEE International Symposium on High-Performance Computer Architecture (HPCA)}}. IEEE, \bibinfo{pages}{273--286}.
\newblock


\bibitem[Zadeh et~al\mbox{.}(2020)]%
        {zadeh2020gobo}
\bibfield{author}{\bibinfo{person}{Ali~Hadi Zadeh}, \bibinfo{person}{Isak Edo}, \bibinfo{person}{Omar~Mohamed Awad}, {and} \bibinfo{person}{Andreas Moshovos}.} \bibinfo{year}{2020}\natexlab{}.
\newblock \showarticletitle{Gobo: Quantizing attention-based nlp models for low latency and energy efficient inference}. In \bibinfo{booktitle}{\emph{2020 53rd Annual IEEE/ACM International Symposium on Microarchitecture (MICRO)}}. IEEE, \bibinfo{pages}{811--824}.
\newblock


\bibitem[Zhou et~al\mbox{.}(2022)]%
        {zhou2022transpim}
\bibfield{author}{\bibinfo{person}{Minxuan Zhou}, \bibinfo{person}{Weihong Xu}, \bibinfo{person}{Jaeyoung Kang}, {and} \bibinfo{person}{Tajana Rosing}.} \bibinfo{year}{2022}\natexlab{}.
\newblock \showarticletitle{Transpim: A memory-based acceleration via software-hardware co-design for transformer}. In \bibinfo{booktitle}{\emph{2022 IEEE International Symposium on High-Performance Computer Architecture (HPCA)}}. IEEE, \bibinfo{pages}{1071--1085}.
\newblock


\end{thebibliography}

\end{document}